\documentclass[letterpaper, 10 pt, conference]{ieeeconf}
\IEEEoverridecommandlockouts 
\usepackage{graphicx} 
\usepackage{cite}
\usepackage{caption}
\usepackage{amsmath}
\usepackage{amsfonts}
\usepackage{dsfont}
\let\labelindent\relax
\usepackage{enumitem,kantlipsum}
\usepackage{subcaption}
\usepackage{multirow}
\usepackage{color}
\DeclareMathOperator*{\argmax}{argmax}

\title{\LARGE \bf Scalable Adaptive Traffic Light Control Over a Traffic Network Including Turns, Transit Delays, and Blocking}

\author{Yingqing Chen and Christos G. Cassandras
\thanks{Y. Chen and C. G. Cassandras are
with the Division of Systems Engineering and Center for Information and
Systems Engineering, Boston University, Brookline, MA 02446
{\tt\small\{yqchenn;cgc\}@bu.edu}. This work was supported in part by Red Hat Research (Red Hat Collaboratory
at Boston University), by NSF under grants ECCS-1931600, DMS-
1664644, CNS-2149511, and by ARPA-E under grant DE-AR0001282.}
}

\begin{document}
\maketitle
\thispagestyle{empty}
\pagestyle{empty}

\begin{abstract} 
We develop adaptive data-driven traffic light controllers for a grid-like traffic network considering straight, left-turn, and right-turn traffic flows. The analysis incorporates transit delays and blocking effects on vehicle movements between neighboring intersections. 
Using a stochastic hybrid system model with parametric traffic light controllers, we use Infinitesimal Perturbation Analysis (IPA) to derive a data-driven cost gradient estimator with respect to controllable parameters. We then iteratively adjust them through an online gradient-based algorithm to improve performance metrics. By integrating a flexible modeling framework to represent diverse intersection and traffic network configurations with event-driven IPA-based adaptive controllers, we develop a general scalable, adaptive framework for real-time traffic light control in multi-intersection traffic networks.
\end{abstract}

\section{Introduction}
The Traffic Light Control (TLC) problem aims to dynamically optimize traffic light cycles at one or multiple intersections to improve traffic flow and reduce congestion as measured by metrics such as waiting times, vehicle throughput, or traffic backlog.
The single intersection TLC problem has been thoroughly studied using different approaches such as model-based optimization \cite{zhang_traffic_2017}, computational intelligence  \cite{kaur_adaptive_2014},\cite{dong_analysis_2019}, and online optimization methods \cite{fleck_adaptive_2016}. In addition to pure vehicle flows, some prior works \cite{zhu_context-aware_2022}\cite{chen2023adaptive}\cite{zhang_traffic_2017} have incorporated pedestrian flows and their interactions with vehicle traffic in addition to optimizing vehicle movement.

However, the transition from single to coordinated multi-intersection control poses at least four key challenges: 
$(i)$ Different traffic flows (straight, turning, U-turn) propagate through the network, complicating efforts to model network-wide effects; 
$(ii)$ \emph{Scalable} solutions that extend one-dimensional approaches to high-dimensional multi-intersection settings are difficult to obtain; 
$(iii)$ The transit delays experienced by traffic between intersections must be accounted for in order to effectively coordinate an intersection with its downstream counterpart(s); and 
$(iv)$ When the distance between adjacent intersections is relatively short, traffic is often blocked and such blocking effects must be taken into account. Capturing the dynamics of propagating flows, achieving scalable optimization, modeling delays between-intersections, and handling blocking constraints are critical factors to achieve effective multi-intersection traffic light control.
%and many solution algorithms have been proposed.

Early works on multi-intersection traffic signal control focused on simple scenarios. For example, \cite{gershenson_self-organizing_2005} presented three self-organizing methods for responsive signal timing, and \cite{hassin_flow_1996} framed fixed-cycle control as a network synchronization problem, only considering offset costs between adjacent intersections.
Later approaches expanded controllable parameters to develop centralized controllers optimizing coordination across multiple signals. For example, \cite{he_steady-state_2015} proposed a steady-state signal control approach which models the traffic light network as a linear time-varying system, and controls the discharging ratio of internal network links. More recently, \cite{wang_optimizing_2022} developed an adaptive linear quadratic regulator to synchronize a network of intersections. However, it is challenging for these model-based approaches to extend to large networks and accommodate the nonlinearities of flow dynamics and stochastic traffic effects.

To deal with the high computational intensity of such centralized optimization methods, various decomposition approaches have been proposed. In \cite{lin_efficient_2012}, the problem is split into nonlinear non-convex subnetwork optimizations to minimize delays and Model Predictive Control (MPC) is used to coordinate subnetwork controllers. Similarly, \cite{huang_adaptive_2018} considers both deterministic traffic demand from commuters and stochastic demand from infrequent travellers. A two-stage stochastic modeling approach is designed to deal with two kinds of demand sequentially by minimizing the expected total travel time at a traffic equilibrium scenario. 
To address the computational complexity caused by nonlinearities, \cite{le_linear-quadratic_2013} presents a linear prediction model and employs an MPC controller using a Quadratic Program (QP) formulation, which quadratically penalizes the number of vehicles in the network and linearly penalizes the control decision.

Although many methods have been implemented to reduce computational requirements, especially in large traffic networks, \cite{BazzanAnaL.C.2005ADAf} argues that centralized approaches to traffic signal control cannot cope with the increasing complexity of urban traffic networks. Thus, decentralized approaches have emerged. El-Tantawy et al. \cite{el-tantawy_multi-agent_2012} designed a decentralized framework where agents coordinate through joint policy learning. They later improved agent coordination using modular Q-learning with neighbor information \cite{el-tantawy_multiagent_2013}. Modeling duration changing action as a high-dimensional Markov decision process, a deep reinforcement learning framework for traffic light cycle control was developed in \cite{liang_deep_2019}. Adding more features, \cite{wang_stmarl_2022} proposed multi-agent reinforcement learning using a Recurrent Neural Network (RNN) and a Graph Neural Network (GNN) to model spatio-temporal influences. 

Decentralized techniques aim to improve scalability and adaptability compared to centralized control. However, both model-based and learning-based approaches mentioned above have downsides in large networks: Centralized optimization faces the ``curse of dimensionality'' as complexity grows, while decentralized reinforcement learning shifts the computational load to extensive training with large historical datasets, especially for large-scale urban traffic networks. Moreover, many models lack the flexibility to adapt to changing traffic conditions, since they are designed or trained for specific traffic patterns and road layouts. Additional effort is needed to adjust models for new scenarios that may occur. Therefore, key unmet challenges remain centered around the difficulty of balancing optimality, scalability, adaptivity, and flexibility. 
%given limited data. 
Open questions persist on coordinating decentralized agents while optimizing system-wide performance under varying demand.

With this motivation, we propose a flexible traffic modeling framework that can incorporate different network topologies. Based on this enhanced network representation, we design scalable adaptive traffic light controllers to optimize signal timings across a multi-intersection structure. Our main contributions are as follows:

$(i)$ We exploit the \emph{scalability} properties of the single-intersection adaptive data-driven TLC approach in \cite{fleck_adaptive_2016},\cite{chen2023adaptive} based on Infinitesimal Perturbation Analysis (IPA) used to estimate on line performance gradients with respect to the parameters of a TLC controller. In particular, since IPA-based gradient estimators are entirely \emph{event-driven}, these algorithms scale with the (relatively small) number of events in each system intersection, not the (much larger) state space dimensionality \cite{chen_stochastic_2020},\cite{chen2023adaptive}. Moreover, IPA is independent of any modeling assumptions regarding the stochastic processes characterizing traffic demand and vehicle behavior, driven only by actual observed traffic data similar to other learning-based approaches.
%These gradient estimates are then used to iteratively seek optimal values for GREEN cycles at each intersection. 

$(ii)$ We expand the multi-intersection framework in \cite{chen2023scalable} (limited to straight flows) by \emph{incorporating turning flows} into the traffic model. By accounting for vehicle turns at each intersection, the model can better capture the propagation of congestion effects across the intersection network and provides the opportunity to optimize coordinated signal timing plans.

$(iii)$ We incorporate vehicle \emph{transit delays} into the traffic network. 
Similar to the  stochastic hybrid system model in \cite{chen_stochastic_2020}, the traffic flow at each intersection depends on the incoming flow from an upstream intersection delayed by the transit time between intersections. In this paper, we provide a simpler way relative to \cite{chen_stochastic_2020} that can capture the flow delay process.
With the incorporation of turning flows, this new framework allows us to take advantage of one of the fundamental properties of IPA \cite{cassandras_perturbation_2010} applied to networks, whereby the effect of a parameter perturbation at one traffic light can only propagate to adjacent traffic lights as a result of a limited and easily tractable number of events; this facilitates the estimation of \emph{network-wide performance} gradients with respect to a local parameter at a specific traffic light.

$(iv)$ We expand the multi-intersection traffic model from \cite{chen2023scalable} by incorporating finite road capacities and analyzing \emph{blocking effects}. Blocking occurs when downstream congestion causes vehicle queues to propagate backwards, preventing movement at upstream intersections and potentially creating gridlock across the network.
The key enhancement is to consider blocking effects not just on the current road segment, but also on upstream segments affected so that the coordinated multi-intersection model can minimize congestion propagation.

$(v)$ We propose a flexible traffic modeling framework by abstracting several traffic features and their relationships over network topologies. By focusing on the traffic features and their interactions rather than specific predefined structures, our framework provides flexibility to model heterogeneous road structures, lane configurations, traffic directions, intersection designs, and traffic signal patterns.

The remainder of this paper is organized as follows. In section II, we formulate the TLC problem for a grid-like multi-intersection network and present the stochastic hybrid system modeling framework. Section III details the derivation of the IPA gradient estimators for a network-wide cost function with respect to a controllable parameter vector. The IPA estimators are then incorporated into a gradient-based optimization algorithm. In Section IV, we conduct multiple simulation experiments under different settings and demonstrate the adaptivity and scalability properties of this approach. Finally, we conclude and discuss future work in Section V.

\section{Problem Formulation}\label{sec:Problem Formulation}
%\subsection{Multiple Signalized Intersection Modeling}
\begin{figure}
    \centering
    \includegraphics[width=0.9\columnwidth]{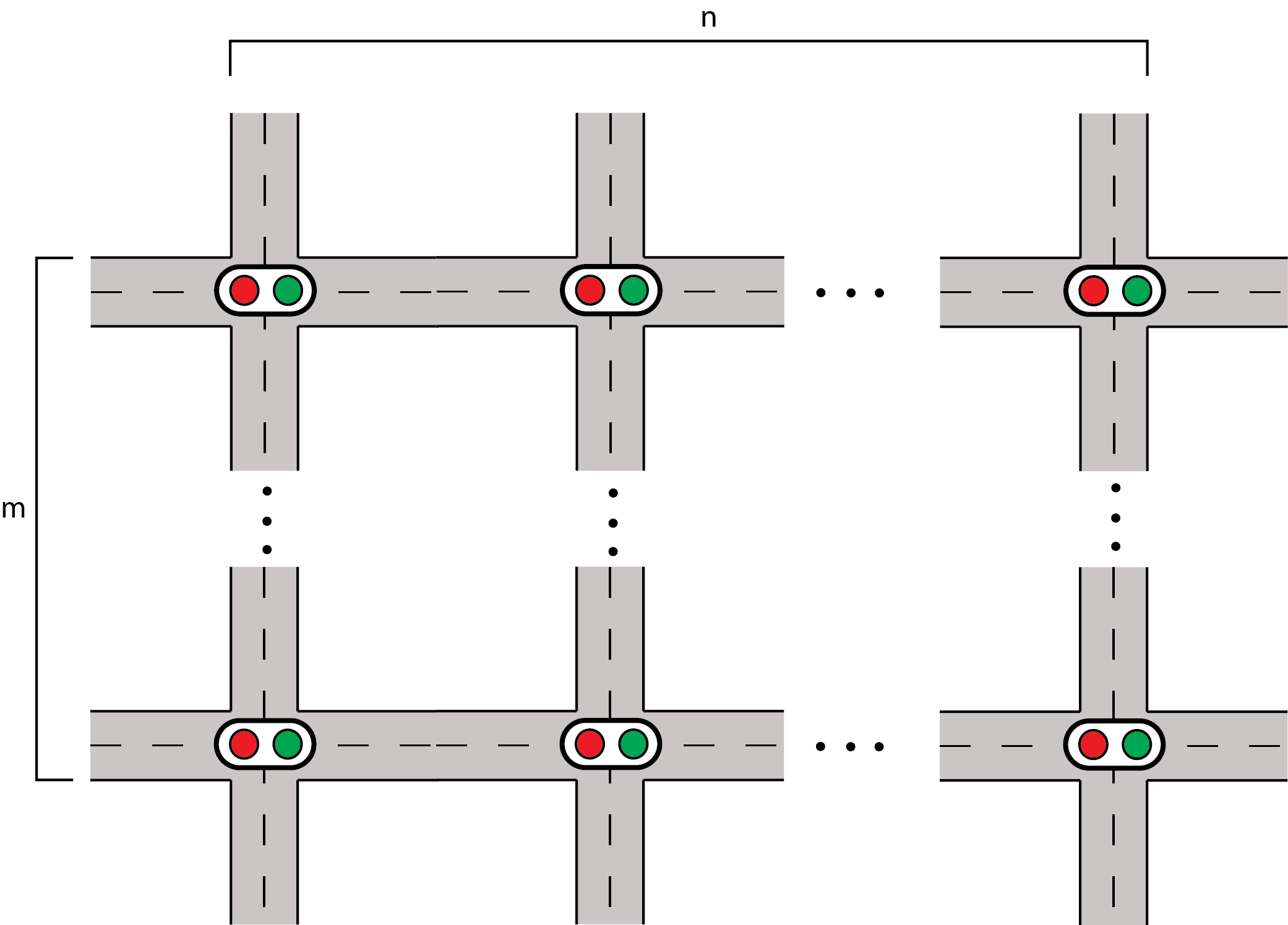}
    \caption{Grid traffic system}
    \label{fig:intersections}
\end{figure}

We consider a traffic system with $m$ parallel arteries, each of which contains $n$ consecutive intersections. A total of $N$ traffic lights are included in this grid-like network, where $N=m*n$, as shown in Fig.\ref{fig:intersections}. 
We assume that the YELLOW cycle is combined with the RED one. The traffic flow towards each intersection $n$ ($n=1,\ldots,N$) can be identified by the orientation of its origin (East, South, West, North) and direction (straight, left, right), and we define each combination of these three attributes as a \emph{movement}. 
For example, movement [n-E-l] denotes the direction for traffic that starts from East and turns left at intersection $n$. We categorize movements into two types: controllable and uncontrollable. Controllable traffic flows can be directed by traffic lights, while uncontrolled traffic is free to move without signals, such as some right-turning vehicle flows.

We can further group movements by their location since the traffic flow for certain movements is confined to specific lanes on the road. Note that several different movements can share a single lane 
%(as a shared lane), 
rather than having dedicated lanes for each movement. We model each lane segment between two adjacent intersections as a \emph{queue} where vehicles stop when the light for the associated movement is RED.  We define the set of all queues at intersection $n$ as $Q_n$. As in actual traffic systems, a traffic light gives right-of-way by showing GREEN to a particular set of controllable movements. After its GREEN cycle time, the light pattern changes to allow the vehicles from the next group of movements to pass through. Therefore, we define a set of movements that are enabled at the same time as a \emph{phase}, and a traffic light cycle for each intersection $n=1,\ldots,N$ contains a predefined sequence of $k$ phases $P_n = (p_{n,1},p_{n,2},\ldots, p_{n,k})$. 
When a phase is enabled, all movements inside that phase are facing a GREEN light with all associated queues such that their movements are consistent in terms of traffic rules.
%The movements within a queue should be included in the same phase so as to ensure that all movements within a queue are consistent in terms of traffic rules. 
Therefore, we denote the corresponding set of enabled queues within phase $p \in P_n$ as $Q_p$, where $Q_p \subset Q_n$.  Note that a single queue could appear in different phases. This indicates that when the GREEN light switches from one phase to the next consecutive phase, some queues may remain enabled if they are included in both phases. Additionally, note that each queue with controllable movements should be enabled at least once during a full signal cycle. This ensures that all movements get a chance to proceed during the cycle.

We now define the corresponding queue content state variable $x_q(t)$ for each queue $q$ in $Q_n$, where $n=1,\ldots, N$, and $x_q(t) \in \mathbb{R}_0^+$. The potential downstream queues for each queue $q$ are determined based on the movements accepted in $q$ and are denoted as $Q^D_q$. Similarly, the set of upstream queues for queue $q$ is denoted as $Q^U_q$. With these set definitions, we capture all the flow connection relationships between queues.

\begin{figure}
    \centering
    \includegraphics[width=0.9\columnwidth]{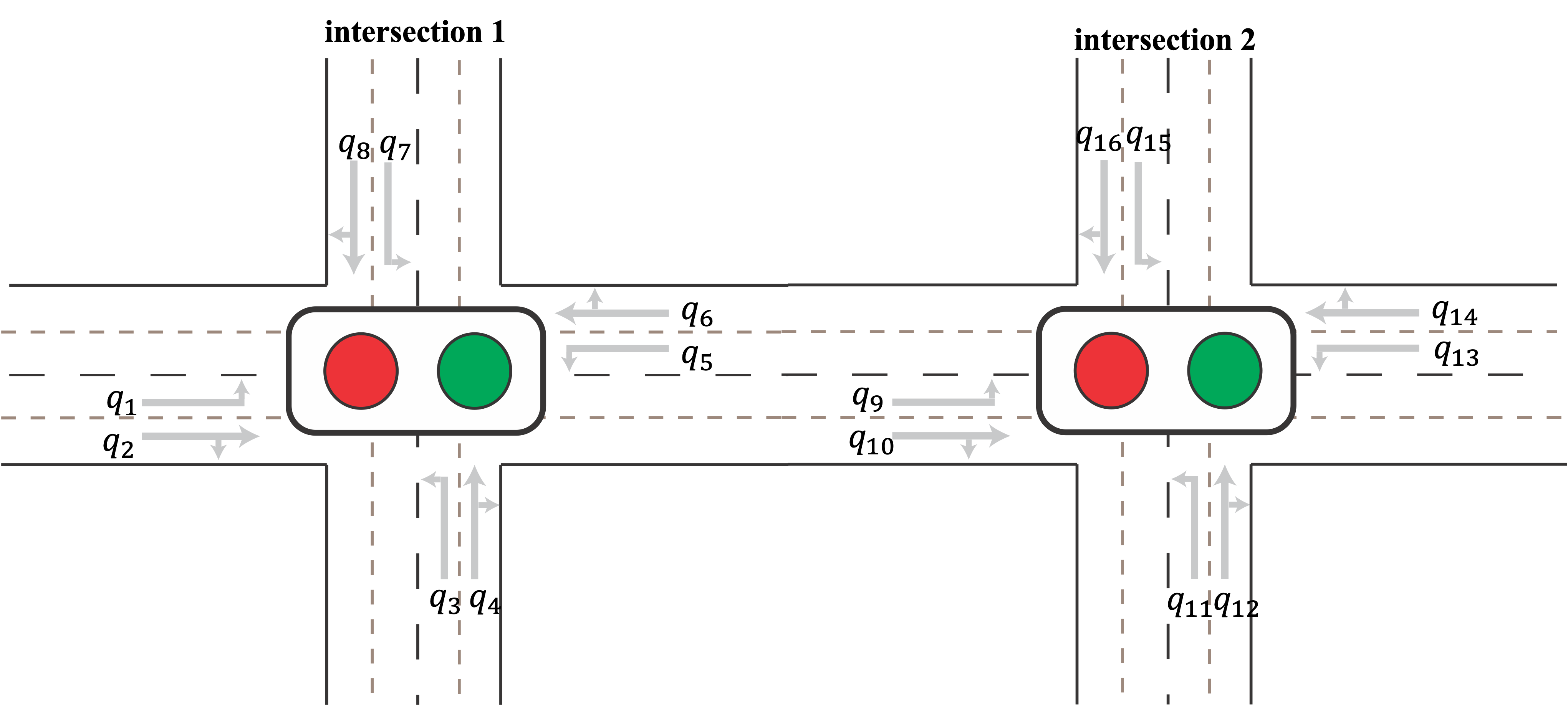}
    \caption{Traffic Structure Demonstration}
    \label{fig:traffic structure}
\end{figure}

As an example, Fig.\ref{fig:traffic structure} shows a portion of a traffic network structure where each road segment contains 2 queues - one for left-turn movements and one for straight and right-turn movements. The two intersections have a similar structure, so we will focus on modeling intersection $1$.  The queue set for intersection $1$ is $Q_1=[q_1,q_2,q_3,q_4, q_5,q_6,q_7,q_8]$. We designate all right-turn movements as uncontrollable, aligning with many real-world traffic regulations. For the remaining movements, we define 4 phases: 
(a) Phase $p_1$ includes straight movement from West [W-s] on queue $q_2$ and straight movement from East [E-s] on queue $q_6$ ($q_2,q_6 \in Q_{p_1}$); 
(b) Phase $p_2$ includes left turns from West [W-l] on $q_1$ and from East [E-l] on $q_5$ ($q_1,q_5 \in Q_{p_2}$); 
(c) Phase $p_3$ includes straight flows from South [S-s] on $q_4$ and from North [N-s] on $q_8$ ($q_4,q_8 \in Q_{p_3}$); and 
(d) Phase $p_4$ includes left turns from South [S-l] on $q_3$ and from North [N-l] on $q_7$ ($q_3,q_7 \in Q_{p_4}$). To connect adjacent intersections, we can specify downstream queue sets under 2-intersection scope, e.g. $Q^D_{q2} = \{q_9,q_{10}\}$, and upstream queue sets, e.g. $Q^U_{q_9} = \{q_2,q_4, q_7\}$. 

With these definitions, we can represent several key features of transportation networks: 
(a) Different acceptable traffic directions (e.g. straight, left-turn or right-turn) are defined by specifying allowable movements; 
(b) Road structures with different numbers of lanes are represented by defining multiple queues for road segments; 
(c) Lane occupancy rules regarding which movements can use which lanes are encoded by assigning sets of allowed movements to each queue; 
(d) Intersection structures like 4-way or 3-way intersections are captured by properly defining the upstream and downstream queue connectivity for each queue based on the intersection geometry; 
(e) Traffic light patterns are specified by defining the corresponding sequence of phases.
In summary, this network representation is flexible enough to model heterogeneous road structures, lane configurations, traffic directions, intersection designs and traffic signal patterns through careful specification of movements, queues, phases and their relationships. This allows the analytical modeling of traffic flows through diverse transportation network topologies.

\textbf{Blocking.} Blocking arises when the road length between two neighboring intersections is not long enough or the green cycle for the in-between queues is too short so that the probability that one or more queues are filled with vehicles is strictly positive. We denote the capacity of queue $q$ as $c_q$ so that when $x_q(t)$ reaches $c_q$ from below, blocking occurs and no more vehicles can join this queue.
%indicating the maximum vehicle number that $q$ can contain. Therefore, when $x_q(t)$ reaches $c_q$ from below, blocking occurs and no more vehicles can join this queue. 
Note that such blocking may not only affect the arrival process at queue $q$, but also halt the departure processes of all upstream queues $q^u \in Q^U_q$, since the vehicles in them are not able to proceed. 
%Note that based on different behavior of drivers tending to join blocking queue (wait inside intersection or not leave the upstream road), the range affected queues is different. We here consider a polite manner and only the upstream queues corresponding to the blocking queue.

\textbf{State Dynamics.} As in prior work, e.g., \cite{fleck_adaptive_2016}, we model
the input and output to each queue as two stochastic flow processes
$\{ \alpha_q(t) \}$ and $\{ \beta_q(t) \}$, where $\alpha_q(t)$ and $\beta_q(t)$ are the \emph{stochastic} instantaneous arrival and departure rate at queue $q$ respectively.
We can then write the queue flow dynamics as
\begin{equation}
\label{eqn:xdynamic}
  \dot x_q(t) =\alpha_q(t)-\beta_q(t), ~~~~~q \in Q_n, n=1,\ldots N
\end{equation}
Unlike a single intersection, where \{$\alpha_q(t)$\} is an exogenous arrival process independent of the state of the system, in the multi-intersection case, the arrival flow process is determined by the departure process of its upstream queues, except when there are no upstream queues, i.e., $q$ is such that $Q^U_q= \emptyset$. The arrival rate $\alpha_q(t)$ for $q \in \{q | Q^U_q \neq \emptyset\}$ will be defined in the sequel so as to properly capture its dependence on $\beta_{q^u}(t), q^u \in Q^U_q$ and include the effect of flow transit delays. On the other hand, 
$\{ \beta_q(t) \}$ depends on the corresponding traffic light control, denoted by $v_q(t)$, 
where $v_q(t)=1$ corresponds to a GREEN phase for queue $q$, and $v_q(t)=0$ to a RED phase, respectively. We set $v_q(t)$ to be a right-continuous function of $t$.
The departure process is also influenced by whether or not the downstream queue is blocked, since a blocked downstream queue prevents vehicles in the corresponding upstream queue from moving, hence $\beta_q(t)=0$. 
Therefore, $\beta_q(t)$ can be expressed as follows:
\begin{align} \label{beta}
  \beta_q(t) =
    \begin{cases}
      h_q(t), & \text{if $x_q(t)>0$, $v_q(t)=1$,}\\ & \text{and $x_{q_d}(t)<c_{q_d} ~~\forall q_d \in Q^D_q$}\\
      \alpha_q(t), & \text{if $x_q(t)=0$, $v_q(t)=1$,}\\
      & \text{and $x_{q_d}(t)<c_{q_d} ~~\forall q_d \in Q^D_q$}\\
      0, & \text{otherwise}
    \end{cases}   
\end{align}
where $q \in Q_n$, $n=1,\ldots,N$ and $h_q(t)$ is the unconstrained departure rate for which appropriate models can be used (see Section IV).

Next, we ensure that the control of a queue is consistent with phase control. 
Given a phase sequence $P_n = (p_{n,1},p_{n,2},\ldots p_{n,k})$, we require that when the GREEN cycle ends for movements in one phase, then the GREEN signals for the next phase movements in the sequence immediately turn on, while the remaining movements face a RED light.  
We denote the control for phase $p$ as $u_p(t)$ for $p \in P_n, n=1,\ldots N$, where $u_p(t)=1$ indicates a GREEN signal for all movements in $p$, otherwise $u_p(t)=0$. We define $u_p(t)$ to be right-continuous in order to accurately represent the control policy defined in the sequel. Moreover, for any intersection $n$, we require $\sum_{p\in P_n}u_p(t)=1$. Therefore, for any $q \in Q_n$, we can always set the control of queue $q \in Q_p$ once the phase $p$ controls are decided:
\begin{align} \label{eqn:u_q vs u_p}
 v_q(t) =
   \begin{cases}
      1, & \text{if $u_p(t)=1$ for some  $p\in P_n$,~$q \in Q_p$}\\
      0, & \text{otherwise}
   \end{cases}   
\end{align}
%\begin{align} \label{eqn:u_q vs u_p}
 % u_q(t) =
  %  \begin{cases}
 %     1, & \text{if $\exists p\in P_n$ such that $u_p(t)=1$ and $q \in Q_p$}\\
 %     0, & \text{otherwise}
 %   \end{cases}   
%\end{align}
%Note that $u_q(t)$ is also right-continuous, same as $u_p$. 
Note that it is possible for a particular queue to be enabled during several consecutive phase cycle.

Finally, we define clock state variables $z_p(t)$ to measure the time since the current enabled phase last switches to $p$. The dynamics of $z_p(t)$ are:
\begin{equation}
\label{eqn:zdynmic}
  \dot z_p(t) =
    \begin{cases}
      1, & \text{if $u_p(t)=1$ }\\
      0, & \text{otherwise} 
    \end{cases}       
\end{equation}
We define $z_p(t)$ to be left-continuous, so that when the enabled phase switches from $p$ to the next phase, $z_p(t)>0$,  $z_p(t^+)=0$, and $u_p(t)=0$ (since $u_p(t)$ is right-continuous).

\textbf{Traffic Light Control.} 
An effective controller design needs to address two issues: 
(i) maintaining a proper balance between allocating a GREEN light to competing queues and 
(ii) preventing the undesired phenomenon where vehicles wait at a RED light while competing queues are empty during their GREEN phase. 
Such “waiting-for-nothing” instances waste the resources of vehicles that wait unnecessarily and can be eliminated through a proper controller design as detailed next. 
Towards these two goals, we design a quasi-dynamic controller such that only partial state information is needed. Such control is based on the ability of current sensors to detect events of interest in the state dynamics above, such as a queue content becoming empty. On the other hand, it may not be possible to detect the exact number of vehicles in a queue (e.g., using cameras); therefore, we assume that this number can be estimated so as to classify a queue content $x_q(t)$ as being empty and either below or above some \emph{queue content threshold}, as well as the time such transitions occur. We define an adjustable queue content threshold, denoted by $s_p$, for each phase $p$. When phase $p$ is enabled, the detector uses the same threshold value $s_p$ for all queues $q \in Q_n$. We also assign a guaranteed minimum GREEN light cycle time $\theta_p^{min}$ and a maximum cycle time $\theta_p^{max}$ for each phase $p$. This is to ensure that traffic light switches are not overly frequent nor can they be excessively long. Thus, a controller for phase $p$ depends on a controllable parameter vector for each phase:
\begin{equation} \label{parametervector}
    \theta_p = [\theta_p^{min}, \theta_p^{max}, s_p]
\end{equation}
for $p \in P_n, n =1,\ldots N$, where $\theta_n^{min}\ge 0$, $\theta_p^{max}\ge \theta_p^{min}$, $s_p\ge 0$. We also define the complete parameter set 
$\Theta=\{\theta : \theta \in \theta_p, ~p \in P_n, ~n=1,\ldots ,N\}$.

\begin{figure}
    \centering
    \includegraphics[width=0.7\columnwidth]{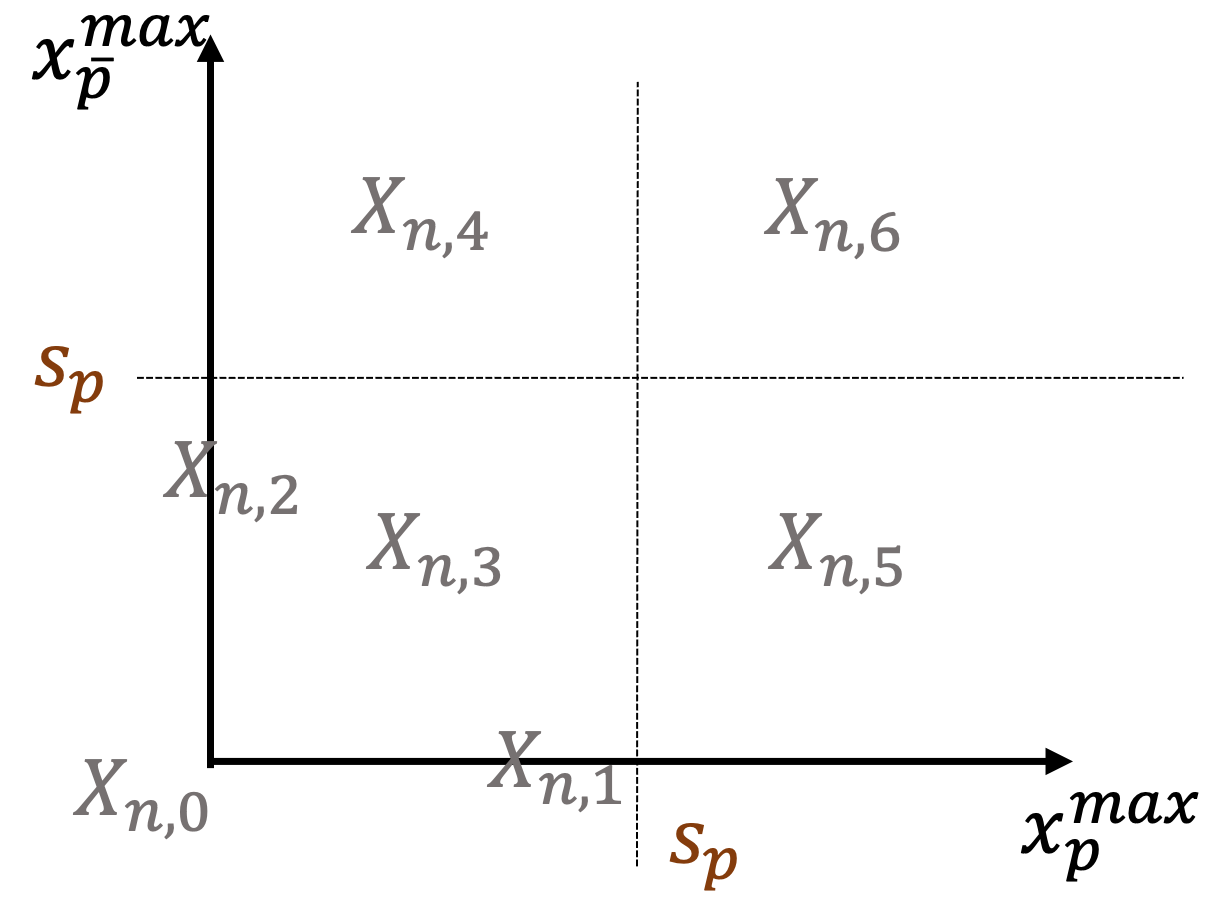}
    \caption{State space representation}
    \label{fig:partition}
\end{figure}

In order to partition the queue content state space of intersection $n$ which is currently under phase $p \in P_n$ ($u_p(t)=1$), we start by defining two auxiliary variables indicating the (instantaneous) largest queue content of any queue in phase $p$ and that of any queue not in phase $p$ as $x_p^{max}(t) = \max_{q \in Q_p} x_q(t)$ and $x_{\bar p}^{max}(t) = \max_{q \in Q_n \backslash Q_p} x_q(t)$. Based on the value of these two variables, we partition the queue content state space into the following regions (as shown in Fig.\ref{fig:partition}):\\
$ X_{n,0}=\{(x_p^{max}, x_{\bar p}^{max}):x_p^{max}(t)=0,  ~x_{\bar p}^{max}(t)=0 \} $\\
$ X_{n,1}=\{(x_p^{max}, x_{\bar p}^{max}):x_p^{max}(t)>0,  ~x_{\bar p}^{max}(t)=0 \} $ \\
$ X_{n,2}=\{(x_p^{max}, x_{\bar p}^{max}):x_p^{max}(t)=0,  ~x_{\bar p}^{max}(t)>0 \} $\\
$X_{n,3}=\{(x_p^{max}, x_{\bar p}^{max}):0<x_p^{max}(t)<s_p, ~0< x_{\bar p}^{max}(t)<s_p \}$\\
$X_{n,4}=\{(x_p^{max}, x_{\bar p}^{max}):0<x_p^{max}(t)<s_p,  ~x_{\bar p}^{max}(t)\geq s_p \}$\\
$X_{n,5}=\{(x_p^{max}, x_{\bar p}^{max}):x_p^{max}(t)\geq s_p, ~0< x_{\bar p}^{max}(t)<s_p \}$\\
$X_{n,6}=\{(x_p^{max}, x_{\bar p}^{max}):x_p^{max}(t)\geq s_p,  ~x_{\bar p}^{max}(t)\geq s_p \}$

Therefore, if intersection $n$ is currently in phase $p \in P_n$, 
%when facing different subset of queue content state space, 
the quasi-dynamic traffic light controller follows the following rules:
\begin{itemize}[wide]

    \item [1.] $(x_p^{max},x_{\bar p}^{max}) \in \{X_{n,1}\}$:
In this case, there is no vehicle flow for any queue in $\{\bar q| \bar q \in Q_n, \bar q \notin Q_p\}$ competing with queue $q$ while queues enabled in the current phase are not empty. In this case, the light for the current phase should remain GREEN:
\begin{equation}
\label{eqn:control rule X1}
  u_p(t) = 1
\end{equation}

\item [2.] $(x_p^{max},x_{\bar p}^{max}) \in \{X_{n,2}\}$:
In this case, there is no vehicle flow for queues enabled in the current phase $p$, while competing queues are not empty. In this case, the current phase should switch to RED immediately:
\begin{equation}
\label{eqn:control rule X2}
  u_p(t) = 0
\end{equation}

\item [3.] $(x_p^{max},x_{\bar p}^{max}) \in \{X_{n,0}, X_{n,3}, X_{n,5},X_{n,6}\}$:
In this case, vehicle traffic for queues in the current phase $p$ is either higher or balanced compared to competing queues. Thus, the current phase should be prioritized and remain GREEN until it reaches its maximum cycle time:
\begin{equation}
\label{eqn:control rule X3}
    u_p(t) = 
    \begin{cases}
      1, & \text{if $z_p(t)\in (0, \theta_p^{max})$}\\
      0, & \text{otherwise}
    \end{cases}  
\end{equation}

\item [4.] $(x_p^{max},x_{\bar p}^{max}) \in \{X_{n,4}\}$:
In this case, there is low vehicle traffic demand for queues in the current phase $p$ and high demand for the other queues, thus the system should switch phases to accommodate the excess demand:
\begin{equation}
\label{eqn:control rule X4}
    u_p(t) = 
    \begin{cases}
      1, & \text{if $z_p(t)\in (0, \theta_p^{min})$}\\
      0, & \text{otherwise}
    \end{cases}  
\end{equation}

\end{itemize}

Note that when the traffic light for movements in the current phase $p$ switches from GREEN to RED, the movements for the next phase in the predefined phase sequence $P_n = (p_{n,1},p_{n,2},\ldots p_{n,k})$ switches from RED to GREEN accordingly.

\textbf{Events.} The state transitions in the model defined through (\ref{beta}), (\ref{eqn:xdynamic}), and (\ref{eqn:zdynmic}) under the controller (\ref{eqn:control rule X1}), (\ref{eqn:control rule X2}), (\ref{eqn:control rule X3}) and (\ref{eqn:control rule X4}) are dictated by ten possible events defined as follows (see also Table \ref{table:events}). 

\emph{Basic events.} For $n=1,\ldots,N, q \in Q_n$ : 
(a) [$x_q\downarrow 0$]: $x_q(t)$ reaches 0 from above,
(b) [$x_q\uparrow 0$]: $x_q(t)$ becomes positive from 0,
(c) [$x_q\downarrow s_p$]: $x_q(t)$ reaches $s_p$ from above,
(d) [$x_q\uparrow s_p$]: $x_q(t)$ reaches $s_p$ from below,
(e) [$x_q\downarrow c_q$]: $x_q(t)$ reaches $c_q$ from above, which indicates the end of a blocking interval,
(f) [$x_q\uparrow c_q$]: $x_q(t)$ reaches $c_q$ from below, which indicates the start of a blocking interval,
(g) [$z_p\uparrow \theta_p^{min}$]: $z_p(t)$ reaches its lower bound,
(h) [$z_p\uparrow \theta_p^{max}$]: $z_p(t)$ reaches its upper bound,
(i) [$\alpha_q\uparrow 0$]: $\alpha_q$ becomes positive from 0,
(j) [$\alpha_q\downarrow 0$]: $\alpha_q$ reaches 0 from above.

\emph{Light switching events. } It is convenient to define the following light switching events which are induced by basic events according to the control rules. For $n=1,\ldots,N, q \in Q_n$:
\begin{itemize}
    \item $R2G_q$: traffic light for controllable movements in queue $q$ at intersection $n$ switches from RED to GREEN.
    \item $G2R_q$: traffic light for controllable movements in queue $q$ at intersection $n$ switches from GREEN to RED. 
\end{itemize}

% Similar concept can also be applied to phase, where the light switching for queue is triggered by phase switch. However, a phase switch does not necessarily lead to a light switch for those affected queues. For $n=1,\ldots,N, p \in P_n$:
% \begin{itemize}
%     \item $R2G_p$: start of GREEN signals for phase $p$, which means movements from phase $p$ are enabled.
%     \item $G2R_p$: end of GREEN signals for phase $p$, which means movements from phase $p$ stop being enabled.
% \end{itemize}

\subsection{Flow Burst Modeling and Analysis}
A major consideration in studying a multi-intersection system is the role of a \emph{flow burst} which is generated at each intersection and has impact on the downstream intersection. This was first studied in \cite{chen_stochastic_2020} for only two intersections and only straight flows using a complicated sequence of sub-processes. Here, we provide a more direct way to model such flow bursts that easily extends to multiple intersections and includes turning flows.
Specifically, when a RED light switches to GREEN for queue $q$, a new flow burst is generated consisting of all vehicles queued at $q$ that are released. 
%We regard this as a single flow with continuously changing rate. 
The impact this flow burst has on any downstream queue $q^d \in Q^D_q$ is caused by the start and end of this flow generation during each GREEN phase. 

We start by defining event $G_{q,m}$ for intersection $n=1,\ldots N$, $q \in Q_n$, $m=1,2,\ldots$ causing the generation of a flow burst from queue $q$ during its $m$th GREEN cycle. This event is induced at time $t$ by one of two already defined events: (a) $R2G_q$ if $x_q(t)>0$, or (b) the first [$\alpha_q\uparrow 0$] event within the $m$th GREEN cycle if $x_q(t)=0$.

When $G_{q,m}$ occurs, $\beta_q(t)$ increases from zero to positive and the corresponding flow burst will join the downstream queue $q^d \in Q^D_q$ (empty or not) following a time delay, which then results in the downstream arrival rate increasing from 0 to positive, thus creating an interdependence of flow rates between neighboring intersections. 
The delay depends on (a) the average speed of the flow burst, denoted by $f_q$, affected by road quality and driver behavior,
(b) the road length $L_{q^d}$ where downstream queue $q^d$ is located, and 
(c) the average length $l$ of a vehicle (including a safe distance between two vehicles).
Then, the flow process relationship between any two adjacent intersections caused by such delay can be estimated as:
\begin{equation}\label{eqn:flow_process_relationship}
    \alpha_{q^d}(t) = \sum_{q\in Q^U_{q^d}}\gamma_{q,q^d}\beta_q(t-\frac{L_{q^d}-x_{q^d}(t)*l}{f_q})
\end{equation}
where $\gamma_{q,q^d}$ is the ratio of flow from $q$ that is directed to $q_d$. 
Note that $L_{q^d}-x_{q^d}(t)*l \ge 0$ always holds even when blocking happens, which implies that $\alpha_{q^d}(t)$ is always dependent on the upstream output process. 
For convenience, set $\Delta_{q,q^d}(t)=\frac{L_{q^d}-x_{q^d}(t)*l}{f_q}$ to be the delay between queues $q$ and $q^d$ in
(\ref{eqn:flow_process_relationship})
and note that $\Delta_{q,q^d}(t)\in [0,\frac{L_{q^d}}{f_q}]$. Thus, (\ref{eqn:flow_process_relationship}) provides the relationship for coupling the flow rates of two consecutive intersections, which allows us to identify events and their corresponding event time derivatives with respect to the controllable parameters $\theta \in \Theta$, as shown in the next section, without needing to know the value of $\gamma_{q,q^d}$.
We also assume that any two flow bursts generated from the same intersection but different GREEN cycles will not join each other before the first one joins the downstream queue. 

\begin{figure}
    \centering
    \includegraphics[width=0.8\columnwidth]{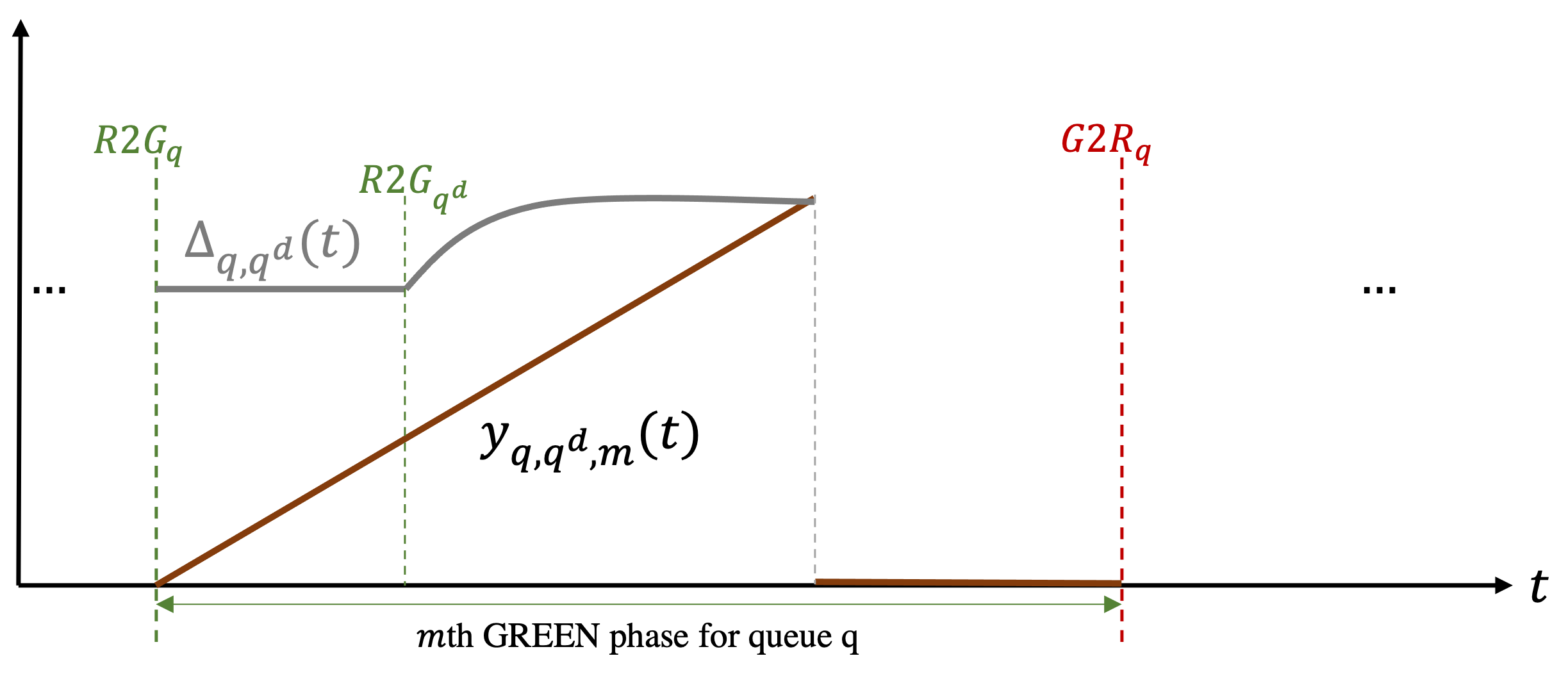}
    \caption{Example for trajectories of $y_{q,q^d,m}(t)$ and $\Delta_{q,q^d}(t)$ }
    \label{fig:y}
\end{figure}

We now define a second clock variable $y_{q,q^d,m}(t) \in \mathbb{R}_0^+$ for intersection $n=1,\ldots N$, $q \in Q_n$, $q^d \in Q^D_q$, $m=1,2,\ldots$, similar to (\ref{eqn:zdynmic}), which denotes the time elapsed since $G_{q,m}$ occurs. Its dynamics are
\begin{equation}\label{eqn:ydynmic}
  \dot y_{q,q^d,m}(t) =
    \begin{cases}
      1, & \text{if $y_{q,q^d,m}(t)\in (0,\Delta_{q,q^d}(t))$}\\
      0, & \text{otherwise}\\
    \end{cases}  
\end{equation}
The clock variable $y_{q,q^d,m}(t)$ has a \emph{single} cycle for each flow burst $m$. When $G_{q,m}$ occurs,
we set $y_{q,q^d,m}(t)=0$ and initialize (\ref{eqn:ydynmic}) so that $y_{q,q^d,m}(t^+)>0$. 
As soon as $y_{q,q^d,m}(t)=\Delta_{q,q^d}(t)$, we set $y_{q,q^d,m}(t^+)=0$. An example is shown in Fig. \ref{fig:y}. 
Note that $y_{q,q^d,m}(t)>0$ is possible for multiple different $m$, i.e., several flow bursts may be active generated from the same intersection but during different GREEN cycles. This may happen when $L_{q^d}$ is large and the flow bursts have yet to reach the downstream queue.
Also note that when $y_{q,q^d,m}(t)=\Delta_{q,q^d}(t)$, this implies that the head of a flow burst generated by event $G_{q,m}$ joins the downstream queue $q^d$ at time $t$. This is the first instant when this flow burst can have an impact on the downstream queue. Therefore, we define it as an event $J_{q,q^d,m}$, which also induces an event [$\alpha_{q^d}\uparrow 0$], i.e., the input flow for queue $q^d$ becomes positive again.

Similarly, in order to model the end of a flow burst, we define an event $G_{q,m}^e$, which indicates the end of a flow burst generation from queue $q$ during the $m$th GREEN cycle. This can be induced by (a) $G2R_q$ if either $\alpha_q(t)>0$ or $x_q(t)>0$, or (b) the last [$\alpha_q\downarrow0$] event inside the $m$th GREEN cycle if $x_q(t)=0$. 
This leads to the definition of another clock state variable $r_{q,q^d,m}(t) \in \mathbb{R}_0^+$ for intersection $n$ where $q\in Q_n, q^d\in Q^D_q$ and $m=1,2,\ldots$, which measures the time elapsed since $G_{q,m}^e$ occurs. Its dynamics are
\begin{equation}\label{eqn:rdynmic}
  \dot r_{q,q^d,m}(t) =
    \begin{cases}
      1, & \text{if $r_{q,q^d,m}(t)\in(0,\Delta_{q,q^d}(t))$}\\
      0, & \text{otherwise}\\
    \end{cases}  
\end{equation}
When $G_{q,m}^e$ occurs, we set $r_{q,q^d,m}(t)=0$ and initialize (\ref{eqn:rdynmic}) so that
$r_{q,q^d,m}(t^+)>0$. As soon as $r_{q,q^d,m}(t)=\Delta_{q,q^d}(t)$, we set $r_{q,q^d,m}(t^+)=0$. 
Similar to $y_{q,q^d,m}(t)$, it is possible that $r_{q,q^d,m}(t)>0$ for multiple different $m$.
Note that when $r_{q,q^d,m}(t)=\Delta_{q,q^d}(t)$, this implies that the flow burst generated from queue $q$ at the $m$th GREEN cycle ceases to have any impact on the downstream queue $q^d$. Therefore, we define it as an event $J_{q,q^d,m}^e$, which also induces an event [$\alpha_{q^d} \downarrow 0$].
Since we consider a single flow burst generated by the same GREEN phase, any [$\alpha_{q^d}\uparrow0$] or [$\alpha_{q^d}\downarrow0$] between $J_{q,q^d,m}$ and $J_{q,q^d,m}^e$ is exogenous and has no impact on the downstream queue state.

In summary, the generation and impact of a flow burst is captured through the four additional \emph{flow burst tracing events} $G_{q,m}$, $J_{q,q^d,m}$, $G_{q,m}^e$, $J_{q,q^d,m}^e$, which facilitate the IPA gradient evaluation in Section \ref{section:ipa}. The full list of events is shown in Table \ref{table:events}. 

% We now have a state vector [$x_q(t), z_q(t), y_{q,q^d}(t), r_{q,q^d}(t)$] for queue $q$. 

\begin{table}[]
\centering
\caption{List of Events}
\resizebox{0.5\textwidth}{!}{%
\begin{tabular}{|l|l|}
\hline
Basic Events & 
 \begin{tabular}[c]{@{}l@{}}{[}$x_q\downarrow 0${]}, {[}$x_q\uparrow 0${]}, {[}$x_q\downarrow s_p${]}, {[}$x_q\uparrow s_p${]}\\ {[}$x_q\downarrow c_q${]}, {[}$x_q\uparrow c_q${]}, {[}$z_p\uparrow \theta_p^{min}${]}\\{[}$z_p\uparrow \theta_p^{max}${]},{[}$\alpha_q\uparrow 0${]}, {[}$\alpha_q\downarrow 0${]}\end{tabular} \\ \hline
% [$x_q\downarrow 0$], [$x_q\uparrow 0$], [$x_q\downarrow s_p$], [$x_q\uparrow s_p$]
% [$z_q\uparrow \theta_q^{min}$],[$z_q\uparrow \theta_q^{max}$],[$\alpha_q\uparrow 0$], [$\alpha_q\downarrow 0$] \\ \hline
    Light Switching Events & $R2G_q$, $G2R_q$ \\ \hline
    Flow Burst Tracing Events & $G_{q,m}$, $J_{q,q^d,m}$, $G_{q,m}^e$, $J_{q,q^d,m}^e$ \\ \hline
\end{tabular}%
}
\label{table:events}
\end{table}

% The head of a flow burst joining the tail of a downstream queue (empty or not) results in the arrival rate for that queue increasing from 0 to positive. Similarly, the tail of this flow burst joining the downstream queue would result in its arrival rate decreasing from positive to 0. In this setting, any event between the time of two joining time related to the upstream flow is exogenous and has no impact on downstream queue state. This model based on ``joining events'' is detailed in what follows.

The multi traffic light intersection system can be viewed as a hybrid system in which the time-driven dynamics are
given by (\ref{eqn:xdynamic}), (\ref{eqn:zdynmic}), (\ref{eqn:ydynmic}), (\ref{eqn:rdynmic}) and (\ref{beta}), while the event-driven dynamics are dictated by the basic events in Table \ref{table:events}; these induce associated light switching and flow burst tracing events (defined for convenience). Although the dynamics are based on knowledge of the instantaneous flow processes $\{ \alpha_q(t) \}$ and $\{ \beta_q(t) \}$, we will show that the IPA-based adaptive controller we design \emph{does not require such knowledge} and depends only on estimating some rates in the vicinity of certain critical observable events.

\subsection{TLC Optimization Problem}
\begin{figure}
    \centering
    \includegraphics[width=0.95\columnwidth]{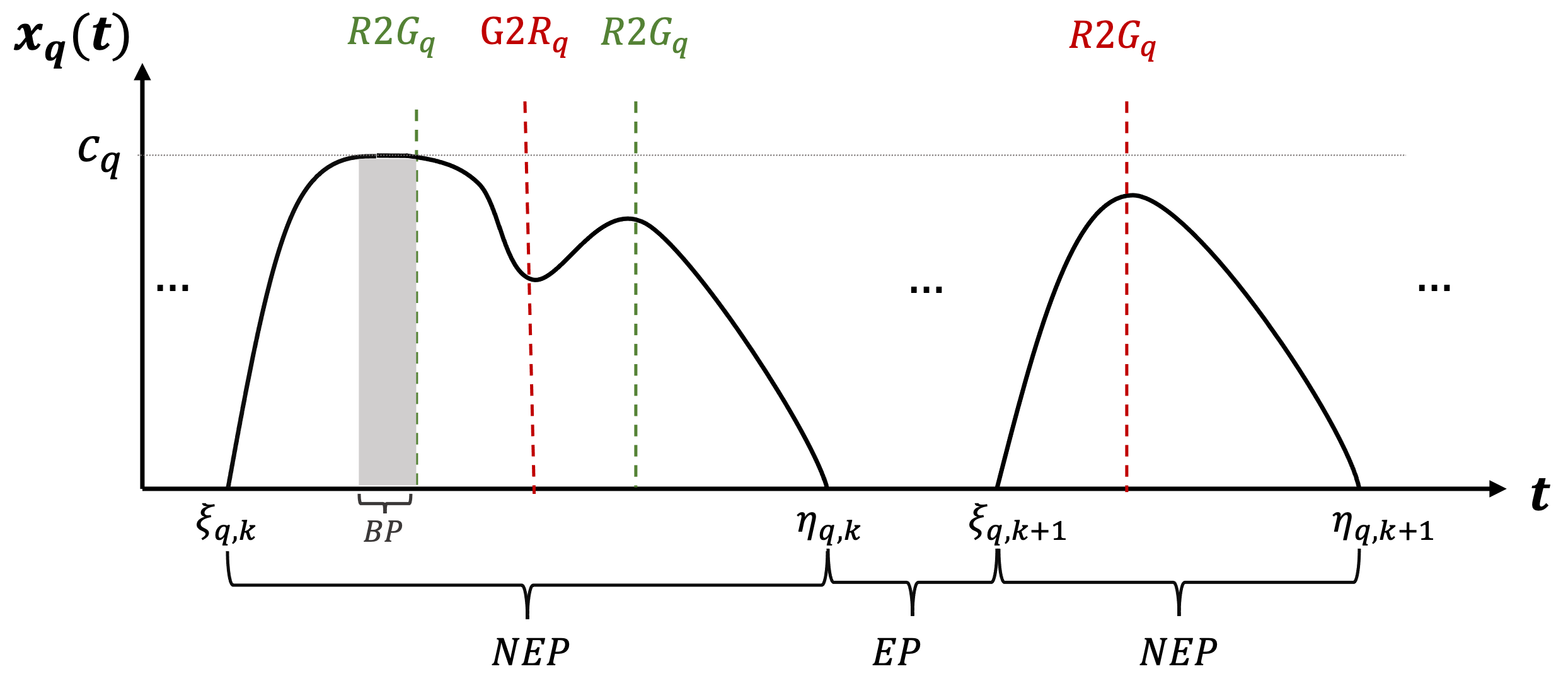}
    \caption{Typical sample path of a traffic queue}
    \label{fig:sample path}
\end{figure}

With the parameterized controller defined 
through $X_{n,0},\ldots,X_{n,6}$ and (\ref{eqn:control rule X1})-(\ref{eqn:control rule X4}), our aim is to optimize a performance metric for the intersection operation with respect to the controllable parameters that comprise $\Theta$ defined through (\ref{parametervector}). 
We choose our performance metric to be the weighted mean of all queue lengths over a fixed time interval $[0,T]$:
\begin{equation} \label{Lfunction}
    L(\Theta;  x(0),z(0), T) = \frac{1}{T}\sum_{n=1}^{N}\sum_{q\in Q_n}\int_{0}^{T} \omega_q x_q(t;\Theta) \,dt 
\end{equation}
where $\omega_q$ is a weight associated with queue $q$.  In order to focus on the structure of a typical sample path of the hybrid system, observe that the sample path of any flow queue content $\{x_q(t)\}$ consists of alternating \emph{Non-empty Periods} (NEPs) and \emph{Empty Periods} (EPs), which correspond to time intervals when $x_q(t)>0$ and $x_q(t)=0$ respectively, as shown in Fig. \ref{fig:sample path}. We define two additional events: $S_q$ for starting NEPs 
and $E_q$ for ending them, both induced by basic events defined earlier. 
Moreover, we denote the $k$th NEP of queue $q$ by $[\xi_{q,k}, \eta_{q,k})$ where $\xi_{q,k}$, $\eta_{q,k}$ are the occurrence times of the $k$th $S_q$ event and $k$th $E_q$ event respectively. Inside NEPs, we further define \emph{Blocking Periods} (BPs) when $x_q(t)=c_q$, indicating blocking happens. 
Since $x_q(t)=0$ during EPs of queue $q$, the sample function $L(\Theta;  x(0),z(0), T)$ in (\ref{Lfunction}) can be rewritten as
\begin{equation}
\label{eqn:L}
    L(\Theta; x(0),z(0), T) = \frac{1}{T}\sum_{n=1}^{N}\sum_{q\in Q_n}\sum_{k=1}^{K_q}\int_{\xi_{q,k}}^{\eta_{q,k}} \omega_q x_q(t;\Theta) \,dt 
\end{equation}
where $K_q$ is the (random) total number of NEPs during the sample path of queue $q$ over $[0,T]$.

Thus, our goal is to determine $\Theta$ that minimizes the expected weighted mean queue length:
\begin{equation}
    J(\Theta; x(0),z(0), T) = E[L(\Theta;  x(0),z(0), T)]
\end{equation}
We note that it is not possible to derive a closed-form expression of $J(\Theta; x(0), z(0), T)$ even if we had full knowledge of the processes $\{ \alpha_q(t) \}$ and $\{ \beta_q(t) \}$. Therefore, a closed-form expression for the gradient $\nabla J(\Theta)$ is also infeasible. 
The role of IPA is to obtain an \emph{unbiased} estimate of $\nabla J(\Theta)$ based on the sample function gradient $\nabla L(\Theta)$ which can be evaluated based only on data directly observable along a single sample path such as Fig. \ref{fig:sample path}, as will be shown in the next section. The unbiasedness of $\nabla L(\Theta)$ is ensured
under mild conditions on $L(\Theta)$ (see \cite{cassandras_perturbation_2010}) and assuming that each $\alpha_q(t)$ is piecewise continuously differentiable in $t$ w.p. 1. In particular, we emphasize that no explicit knowledge of $\alpha_q(t)$ is necessary to estimate $\nabla J(\Theta)$ through $\nabla L(\Theta)$.

We can now invoke a gradient-based algorithm of the form
\begin{equation} \label{gradientopt}
    \theta_{p,i,l+1} = \theta_{p,i,l}-\rho_l \big[\frac{dJ}{d\theta_{p,i,l}}\big]_{IPA}
\end{equation}
where $\theta_{p,i,l}$ is the $i$th parameter of $\theta_p$ at the $l$th iteration ($i\in \{1,2,3\}$), $\rho_l$ is the stepsize at the $l$th iteration, and $[\frac{dJ}{d\theta_{p,i,l}}]_{IPA}$ is the IPA estimator of $\frac{dJ}{d\theta_{p,i,l}}$, which will be derived in the next section.

\section{Infinitesimal Perturbation Analysis} \label{section:ipa}

We begin with a brief review of the IPA framework in \cite{cassandras_perturbation_2010}. Consider a sample path over $[0,T]$ and denote the occurrence time of the $k$th event (of any type) by $\tau_k$. Let $x'(\theta, t)$, $\tau'_k(\theta
)$ be the derivatives of $x(\theta, t)$, $\tau_k(\theta)$ over the scalar controllable parameter of interest $\theta$ respectively. We omit the dependence on $\theta$ for ease of notation hereafter. The dynamics of $x(t)$ are fixed over any inter-event interval $ [\tau_k, \tau_{k+1})$, represented by $\dot{x}(t)=f_k(t)$. Then, the state derivative satisfies 
\begin{equation}
\label{eqn:state_derivative_pre}
    \frac{d}{dt}x'(t)=\frac{\partial f_k(t)}{\partial x}x'(t) + \frac{\partial f_k(t)}{\partial \theta}
\end{equation}
with boundary condition (see \cite{cassandras_perturbation_2010}):
\begin{equation}
\label{eqn:state_derivative}
    x'(\tau_k^+) = x'(\tau_k^-) + [f_{k-1}(\tau_k^-)-f_k(\tau_k^+)]\tau_k'
\end{equation}
In order to evaluate (\ref{eqn:state_derivative}), $\tau_k'$ must be determined, which depends on the type of event taking place at $\tau_k$.
For exogenous events (events causing a discrete state transition that is independent of any controllable parameter), we have $\tau_k'=0$.
For endogenous events (events that occur when there exists a continuously differentiable function $g_k$ such that $\tau_k=min\{t>\tau_{k-1}:g_k(x(\theta, t),\theta)=0\}$) with guard condition $g_k = 0$ (see \cite{cassandras_perturbation_2010}):
\begin{equation}
\label{eqn:event_time_derivative}
    \tau_k' = -[\frac{\partial g_k}{\partial x} f_k(\tau_k^-)]^{-1}(\frac{\partial g_k}{\partial \theta}+\frac{\partial g_k}{\partial x}x'(\tau_k^-))
\end{equation}
This framework captures how system states and event times change with respect to controllable parameters. Our goal is to estimate $\nabla J(\Theta)$ through $\nabla L(\Theta)$, and, according to (\ref{eqn:L}), the performance metric expression is a function of event time and system state variables. Thus, we apply the IPA framework to the TLC problem and evaluate how a perturbation in $\theta$ would affect performance metrics.

\subsection{State Derivatives.}\label{sec:state derivative}
We define the derivatives of the state variable $x_q(t)$, $z_q(t)$, $y_{q,q^d,m}(t)$, $r_{q,q^d,m}(t)$ and event times $\tau_k$ with respect to parameter $\Theta_i$ ($i=1,\ldots,|\Theta|$) as follows:
\begin{multline}
     x_{q,i}'\equiv \frac{\partial x_q(t)}{\partial \Theta_i},~ 
     z_{q,i}'\equiv \frac{\partial z_q(t)}{\partial \Theta_i},~
    y_{q,q^d,m,i}'\equiv \frac{\partial y_{q,q^d,m}(t)}{\partial \Theta_i}, \\
    r_{q,q^d,m,i}'\equiv \frac{\partial r_{q,q^d,m}(t)}{\partial \Theta_i},~~ 
   \tau_{k,i}'\equiv \frac{\partial \tau_k}{\partial \Theta_i} 
\end{multline}
For ease of notation, we denote the state dynamics in (\ref{eqn:xdynamic}), (\ref{eqn:zdynmic}), (\ref{eqn:ydynmic}), (\ref{eqn:rdynmic}) over an inter-event interval $[\tau_k, \tau_{k+1})$ as follows:
\begin{multline}
    \dot x_q(t) = f_{q,k}^x(t),~ \dot z_p(t) = f_{p,k}^z(t), \dot y_{q,q^d,m}(t) = f_{q,q^d,m,k}^y(t),\\\dot r_{q,q^d,m}(t) = f_{q,q^d,m,k}^r(t), q \in Q_n, q^d\in Q_q^D, ~n=1,\ldots,N
\end{multline}
Combining the dynamics in (\ref{eqn:xdynamic}), (\ref{eqn:zdynmic}), (\ref{eqn:ydynmic}) and (\ref{eqn:rdynmic}) with (\ref{eqn:state_derivative_pre}), similar to the analysis in \cite{fleck_adaptive_2016} we can easily conclude that the state derivative of any queue is unaffected within any inter-event time interval, i.e., for $t\in[\tau_k,\tau_{k+1})$:
\begin{multline}\label{eqn:derivative within mode}
    x_{q,i}'(t)=x_{q,i}'(\tau_k^+),\  
    z_{q,i}'(t)=z_{q,i}'(\tau_k^+),\\ 
    y'_{q,q^d,m,i}(t)=y'_{q,q^d,m,i}(\tau_k^+),\  
    r'_{q,q^d,m,i}(t)=r'_{q,q^d,m,i}(\tau_k^+)
\end{multline}
Next, for any discrete event time $\tau_k$, we evaluate queue content derivatives for any possible event occurring to start/end an EP/NEP or within any EP/NEP, and for any controllable parameter $\Theta_i$ ($i=1,\ldots,|\Theta|$):
\begin{itemize}
    \item [1)] Event inside EP: Since $x_q(t)=0$ throughout the whole EP, it immediately follows that
    \begin{equation}\label{eqn:state_der_inside_EP}
        x_{q,i}'(\tau_k^+)=0
    \end{equation}
    
    \item[2)] Event starting EP ($E_q$): This is induced by the basic event [$x_q\downarrow 0$]. The state dynamics change from $f_{q,k-1}^x(\tau_k^-)=\alpha_q(\tau_k^-)-h_q(\tau_k^-)$ to $f_{q,k}^x(\tau_k^+)=\alpha_q(\tau_k^+)-\alpha_q(\tau_k^+)=0$.
    Then, from (\ref{eqn:state_derivative}),
    \begin{equation}
    \label{eqn:state_der_En}
        x_{q,i}'(\tau_k^+)=x_{q,i}'(\tau_k^-)+(\alpha_q(\tau_k^-)-h_q(\tau_k^-))\tau_{k,i}'
    \end{equation}
    
    \item[3)] Event starting NEP ($S_q$) from EP: This can be induced in three possible ways:
    \begin{itemize}
        \item [3.1)]$S_q$ induced by light switching to RED ($G2R_q$) when $\alpha_q(\tau_k)>0$. We have $f_{q,k-1}^x(\tau_k^-)=0$ and $f_{q,k}^x(\tau_k^+)=\alpha_q(\tau_k^+)$. Based on (\ref{eqn:state_der_inside_EP}) and (\ref{eqn:state_derivative}):
        \begin{equation}\label{eqn:state_der_Sn_light_switch}
            x_{q,i}'(\tau_k^+)=-\alpha_q(\tau_k^+)\tau_{k,i}' 
        \end{equation}
        
        \item [3.2)]$S_q$ induced by $J_{q^u,q,m}$ where $q^u \in Q_q^U$. The state dynamics are $f_{q,k-1}^x(\tau_k^-)=0$, $f_{q,k}^x(\tau_k^+)=\alpha_q(\tau_k^+)-\beta_q(\tau_k^+)$, where $\beta_q(\tau_k)$ is defined in (\ref{beta}) and $\alpha_q(\tau_k^+)-\beta_q(\tau_k^+)>0$ in order to induce $S_q$. Based on (\ref{eqn:state_derivative}) we get
        \begin{equation}\label{eqn:state_der_Sn_alpha}
            x_{q,i}'(\tau_k^+)= (\beta_q(\tau_k^+)-\alpha_q(\tau_k^+))\tau_{k,i}'
        \end{equation}
        
        \item [3.3)] $S_q$ induced by an exogenous change in $\alpha_q(\tau_k)$ when $Q^U_q = \emptyset$. In this case, $\tau_{k,i}'=0$, so that we have
            \begin{equation}
        \label{eqn:state_der_Sn_exo}
            x_{q,i}'(\tau_k^+) = x_{q,i}'(\tau_k^-)=0
        \end{equation}
    \end{itemize}

    \item[4)] Event inside NEP. The following are all possible cases:
    \begin{itemize}
        \item [4.1)] $G2R_q$: the state dynamics are $f_{q,k-1}^x(\tau_k^-)=\alpha_q(\tau_k^-)-h_q(\tau_k^-)$ and $f_{q,k}^x(\tau_k^+)=\alpha_q(\tau_k^+)$. Note $\alpha_q(t)$ is continuous except when joining events happen at $q$. Therefore,
        \begin{equation}\label{eqn:state_der_NEP_G2R}
            x_{q,i}'(\tau_k^+)=x_{q,i}'(\tau_k^-)-h_q(\tau_k^-)\tau'_{k,i}
        \end{equation}
            
        \item[4.2)]$R2G_q$: the state dynamics are $f_{q,k-1}^x(\tau_k^-)=\alpha_q(\tau_k^-)$ and $f_{q,k}^x(\tau_k^+)=\alpha_q(\tau_k^+)-h_q(\tau_k^+)$. Therefore,
      \begin{equation}\label{eqn:state_der_NEP_R2G}
            x_{q,i}'(\tau_k^+)=x_{q,i}'(\tau_k^-)+h_q(\tau_k^+)\tau'_{k,i}
        \end{equation}
    
        \item [4.3)]$J_{q^u,q,m}$ with $q^u \in Q^U_q$: the state dynamics are: $f_{q,k-1}^x(\tau_k^-)=-\beta_q(\tau_k^-)$, $f_{q,k}^x(\tau_k^+)=\alpha_q(\tau_k^+)-\beta_q(\tau_k^+)$ where $\beta_q(\tau_k)$ follows (\ref{beta}), and it is continuous except when a light switching event happens at $q$. Thus,
        \begin{equation}\label{eqn:state_der_NEP_J0}
            x_{q,i}'(\tau_k^+)=
                x_{q,i}'(\tau_k^-)-\alpha_q(\tau_k^+)\tau_{k,i}'
        \end{equation}
    
        \item [4.4)] $J_{q^u,q,m}^e$ with $q^u \in Q^U_q$: similarly, the state dynamics are: $f_{q,k-1}^x(\tau_k^-)=\alpha_q(\tau_k^-)-\beta_q(\tau_k^-)$, $f_{q,k}^x(\tau_k^+)=-\beta_q(\tau_k^+)$. Therefore, 
        \begin{equation}\label{eqn:state_der_NEP_J1}
            x_{q,i}'(\tau_k^+)=
            x_{q,i}'(\tau_k^-)+\alpha_q(\tau_k^-)\tau_{k,i}'
        \end{equation}

        \item[4.5)] Other exogenous events. Those events would not affect state derivatives, so that:
        \begin{equation}\label{eqn:state_der_NEP_other}
                x_{q,i}'(\tau_k^+)={x_{q,i}}'(\tau_k^-)
        \end{equation}
        
    \end{itemize}

    \item[5)]\label{item:starting BP} Event starting BP. Blocking is triggered by [$x_q\uparrow c_q$].  When queue $q$ starts to block, in addition to the current state $x_q(t)$, states from all upstream queues $q^u \in Q^U_q$ are also affected. For queue $q$, the state dynamics change from $f_{q,k-1}^x(\tau_k^-)=\alpha_q(\tau_k^-)-\beta_q(\tau_k^-)$ to $f_{q,k}^x(\tau_k^+)=0$. Therefore,
         \begin{equation}\label{eqn:state_der_start_BP_1}
                x_{q,i}'(\tau_k^+)=
            x_{q,i}'(\tau_k^-)+(\alpha_q(\tau_k^-)-\beta_q(\tau_k^-))\tau_{k,i}'
        \end{equation}
    For upstream queue $q^u \in Q^U_q$, the state dynamics change from $f_{q^u,k-1}^x(\tau_k^-)=\alpha_{q^u}(\tau_k^-)-\beta_{q^u}(\tau_k^-)$ to $f_{q^u,k}^x(\tau_k^+)=\alpha_{q^u}(\tau_k^+)$. Therefore,
        \begin{equation}\label{eqn:state_der_start_BP_2}
                x_{q^u,i}'(\tau_k^+)=
            x_{q^u,i}'(\tau_k^-)-\beta_{q^u}(\tau_k^-)\tau_{k,i}'
        \end{equation}

     \item[6)] Event inside BP. Since the state dynamics do not change during blocking, the state derivatives do not change:
        \begin{equation}\label{eqn:state_der_inside_BP}
                x_{q,i}'(\tau_k^+)=x_{q,i}'(\tau_k^-)
        \end{equation}

     \item[7)] Event starting NEP from BP. This is triggered by [$x_q\downarrow c_q$], and, similar to case 5, it affects the state of all upstream queues. For queue $q$, the state dynamics are: $f_{q,k-1}^x(\tau_k^-)=0$, $f_{q,k}^x(\tau_k^+)=\alpha_{q}(\tau_k^+)-\beta_{q}(\tau_k^+)$. Therefore:
     \begin{equation}\label{eqn:state_der_end_BP_1}
                x_{q,i}'(\tau_k^+)=
            x_{q,i}'(\tau_k^-)+(\beta_q(\tau_k^+) -\alpha_q(\tau_k^+))\tau_{k,i}'
        \end{equation}
        
    For upstream queue $q^u \in Q^U_q$, the state dynamics are: $f_{q^u,k-1}^x(\tau_k^-)=\alpha_{q^u}(\tau_k^-)$ to $f_{q^u,k}^x(\tau_k^+)=\alpha_{q^u}(\tau_k^+) - \beta_{q^u}(\tau_k^+)$. Therefore,
        \begin{equation}\label{eqn:state_der_end_BP_2}
                x_{q^u,i}'(\tau_k^+)=
            x_{q^u,i}'(\tau_k^-)+\beta_{q^u}(\tau_k^+)\tau_{k,i}'
        \end{equation}

\end{itemize}

Observe that whenever $Q^U_q \neq \emptyset$, the value of $\alpha_q$ is given by (\ref{eqn:flow_process_relationship}), thus capturing the interdependence of queue content derivatives between adjacent queues $q^u,q$.
Also note that most of the queue content derivative expressions involve the event time derivative $\tau_{k,i}'$.
Therefore, to complete our analysis we need to derive these expressions by applying  
(\ref{eqn:event_time_derivative}) as shown next and use them in (\ref{eqn:state_der_En}), (\ref{eqn:state_der_Sn_light_switch}), (\ref{eqn:state_der_Sn_alpha}), (\ref{eqn:state_der_NEP_G2R}), (\ref{eqn:state_der_NEP_R2G}), (\ref{eqn:state_der_NEP_J0}), (\ref{eqn:state_der_NEP_J1}), (\ref{eqn:state_der_start_BP_1}), (\ref{eqn:state_der_start_BP_2}), (\ref{eqn:state_der_end_BP_1}) and (\ref{eqn:state_der_end_BP_2}).
        
\subsection{Event Time Derivatives}\label{sec:event time derivatives}
In this section, we derive the event time derivatives with respect to each of the controllable parameters $\Theta_i$, $i=1,\ldots,|\Theta|$ as formulated in (\ref{parametervector}). For queue $q \in Q_n$ within intersection $n=1,\ldots, N$:

\begin{itemize}
    \item [1)] Event $E_q$ occurs at $\tau_k$. This is induced by [$x_q\downarrow 0$] so that the guard condition is $g_k = x_q -0=0$, which gives $\frac{\partial g_k}{\partial x_q}=1$, $\frac{\partial g_k}{\partial \Theta_i}=0$. The dynamics are $f_{q,k-1}^x(\tau_k^-)=\alpha_q(\tau_k^-)-h_q(\tau_k^-)$, $f_{q,k}^x(\tau_k^+)=0$. Then, it follows from (\ref{eqn:event_time_derivative}) that:
    \begin{equation}\label{eqn:event_der_En}
    \tau_{k,i}'= \frac{-{x_{q,i}}'(\tau_k^-)}{\alpha_q(\tau_k^-)-h_q(\tau_k^-)} 
    \end{equation}
    This completes equation (\ref{eqn:state_der_En}) and yields ${x_{q,i}}'(\tau_k^+)=0$ for the state derivative in case 2.

    \item[2)] Event $G2R_q$ occurs at $\tau_k$. This happens when there is a phase switch and queue $q$ is part of the previous enabled phase $p$ ($q\in Q_p$) but not part of the next enabled phase $p'$ ($q\notin Q_{p'}$). Such a phase switch can be triggered by the following basic events:

    \begin{itemize}
        \item [2.1)] [$z_p\uparrow \theta_p^{max}$] while $x_{\bar p}^{max}(\tau_k) >0$. The guard condition is $g_k=z_p-\theta_p^{max}$ with $\frac{\partial g_k}{\partial z_p}=1$ and $\frac{\partial g_k}{\partial \theta_p^{max}}=-1$, and all other partial derivatives equals to zero. We also have $f_{p,k}^z(\tau_k^-)= 1$. Similar to the analysis in \cite{fleck_adaptive_2016} we get:
    \begin{equation}\label{eqn:event_der_zp_uparrow_theta_max}
        \tau_{k,i}'= \tau_{k_s,i}'+\mathds{1}_{\Theta_i=\theta_p^{max}}
    \end{equation}
    where $k_s$ is the index of the last light switching event ($R2G_q$) prior to event $k$ and 
    $\mathds{1}_{\cdot}$ is the indicator function. This completes equations (\ref{eqn:state_der_Sn_light_switch}) and (\ref{eqn:state_der_NEP_G2R}).

     \item [2.2)] [$z_p\uparrow \theta_p^{min}$] while $x_p^{max}(\tau_k)<s_p$ and$x_{\bar p}^{max}(t)>s_p$. The guard condition is $g_k=z_p-\theta_p^{min}$. Similar to the last case, (\ref{eqn:event_time_derivative}) can be reduced to:
    \begin{equation}\label{eqn:event_der_zp_uparrow_theta_min}
        \tau_{k,i}'= {\tau_{k_s,i}}'+\mathds{1}_{\Theta_i=\theta_p^{min}}
    \end{equation}

    \item [2.3)] [$x_p^{max} \downarrow s_p$] while $z_p \ge \theta_p^{min}$ and $x_{\bar p}^{max}(\tau_k) \ge s_p$. 
    The guard condition is $g_k=x_{q^*}(\tau_k)-s_p$ with $\frac{\partial g_k}{\partial x_{q^*}}=1$ and $\frac{\partial g_k}{\partial s_p}=-1$, and all other partial derivatives equal to zero, where $q^*=\argmax_{q\in Q_p} x_q(\tau_k)$. Due to light switching, the state dynamics change from $f_{q,k-1}^x(\tau_k^-)=\alpha_q(\tau_k^-)-\beta_q(\tau_k^-)$ to $f_{q,k}^x(\tau_k^+)=\alpha_q(\tau_k^+)$. As a result, (\ref{eqn:event_time_derivative}) reduces to:
    \begin{equation}\label{eqn:event_der_xp_downarrow_sp}
        \tau_{k,i}'= \frac{\mathds{1}_{\Theta_i=s_p}-x'_{q^*,i}(\tau_k^-)}{\alpha_{q^*}(\tau_k^-)-\beta_{q^*}(\tau_k^-)}
    \end{equation}

    \item [2.4)] [$x_{\bar p}^{max} \uparrow s_p$] while $z_p \ge \theta_p^{min}$ and $x_{p}^{max}(\tau_k) < s_p$. 
    The guard condition is $g_k=x_{\bar q^*}(\tau_k)-s_p$ with $\frac{\partial g_k}{\partial x_{\bar q^*}}=1$ and $\frac{\partial g_k}{\partial s_p}=-1$, and all other partial derivatives equal to zero, where $\bar q^*=\argmax_{\bar q\in Q_n\backslash Q_p} x_{\bar q}(\tau_k)$. Similar to the previous case: 
    \begin{equation}\label{eqn:event_der_xpbar_uparrow_sp}
        \tau_{k,i}'= \frac{\mathds{1}_{\Theta_i=s_p}-x'_{\bar q^*,i}(\tau_k^-)}{\alpha_{\bar q^*}(\tau_k^-)}
    \end{equation}
    
    \end{itemize}
    
    When it comes to event $R2G_q$, note that this  is triggered under similar conditions to those of $G2R_q$. In particular, $R2G_{q}$ occurs at phase switching times when queue $q$ is not part of the previous enabled phase $p$ ($q\notin Q_p$) but part of the next enabled phase $p'$ ($q\in Q_{p'}$). Thus, we omit its analysis here. The combined four cases of these events together complete the state derivative calculation of (\ref{eqn:state_der_Sn_light_switch}), (\ref{eqn:state_der_NEP_R2G}), and (\ref{eqn:state_der_NEP_G2R}).

    \item [3)]Event $J_{q^u,q,m}$ occurs at $\tau_k$, where $q^u \in Q^U_q$. This is an endogenous event triggered by  $y_{q^u,q,m}(\tau_k)=\Delta_{q^u,q}(\tau_k)$, so that the associated guard condition is $g_k=y_{q^u,q,m}(\tau_k)-\frac{L_{q}-x_{q}(\tau_k)*l}{f_{q^u}}=0$. Since there are two state variables present in this guard condition, we take derivatives of $g_k$ with respect to parameter $\Theta_i$ first. Since the event time variable $\tau_k$ is directly affected by $\Theta_i$, while state variables $x_q(\tau_k)$ and $y_{q^u,q,m}(\tau_k)$ are both directly and indirectly affected by $\Theta_i$ through $\tau_k$, using the chain rule yields the derivative $g_{k,i}'$ as follows:
    \begin{multline}\label{eqn:gk_der_y}
        g_{k,i}'=y_{q^u,q,m,i}'(\tau_k^-)+f_{q^u,u,m,k-1}^y(\tau_k^-)\tau_{k,i}' + \\ \frac{l}{f_{q^u}}(x_{q,i}'(\tau_k^-)+f_{q,k-1}^x(\tau_k^-)\tau_{k,i}')=0
    \end{multline}
    The last event before $\tau_k$ that would cause the change of state $y_{q^u,q,m}$ is $G_{q^u,q,m}$ (start of flow burst generation). Denoting its occurrence time by $\tau_{k_g}$, we have $f_{q^u,q,m,k_g-1}^y(\tau_{k_g}^-)=0$ and $f_{q^u,q,m,k_g}^y(\tau_{k_g}^+)=1$. Also, since $y_{q^u,q,m}(t)=0$ right before $\tau_{k_g}$, we have $y_{q^u,q,m,i}'(\tau_{k_g}^-)=0$ for all $i=1,\ldots,|\Theta|$.
    Therefore, by applying (\ref{eqn:state_derivative}) and (\ref{eqn:derivative within mode}), we get:
    \begin{equation}\label{eqn:gk_der_y_1}
        \begin{aligned}
        y_{q^u,q,m,i}'(\tau_k^-) &=y_{q^u,q,m,i}'(\tau_{k_g}^+)\\
        &=y_{q^u,q,m,i}'(\tau_{k_g}^-)+(0-1)\tau_{k_g,i}'\\
        &=-\tau_{k_g,i}'
        \end{aligned}
    \end{equation}
    The state dynamics can be calculated from (\ref{eqn:ydynmic}) and (\ref{eqn:xdynamic}) with $\alpha_q(\tau_k^-)=0$:
    \begin{equation}\label{eqn:gk_der_y_2}
        f_{q^u,q,m,k-1}^y(\tau_k^-)=1
    \end{equation}
    \begin{equation}\label{eqn:gk_der_y_3}
        f_{q,k-1}^x(\tau_k^-)= -\beta_q(\tau_k^-)
    \end{equation}
    
    Then, making use of (\ref{eqn:gk_der_y_1}), (\ref{eqn:gk_der_y_2}) and (\ref{eqn:gk_der_y_3}) in (\ref{eqn:gk_der_y}) finally yields:
    \begin{equation}\label{eqn:event_der_J0}
           \tau_{k,i}'=\frac{f_{q^u}}{f_{q^u}-l\beta_q(\tau_k^-)}(\tau_{k_g,i}'-\frac{l}{f_{q^u}}{x_{q,i}}'(\tau_k^-))
    \end{equation}

    where $\tau_{k_g,i}'$ is the event time derivative when $G_{q^u,q,m}$ occurs with [$\beta_{q^u}\uparrow 0$]. This can be induced by: (a) $R2G_{q^u}$ when $x_{q^u}(\tau_{k_g,i})>0$, in which case $\tau_{k_g,i}'$ follows (\ref{eqn:event_der_zp_uparrow_theta_max}), (\ref{eqn:event_der_zp_uparrow_theta_min}), (\ref{eqn:event_der_xp_downarrow_sp}) or (\ref{eqn:event_der_xpbar_uparrow_sp}); (b) $J_{q^{uu},q^{u},\hat{m}}$ when $x_{q^u}(\tau_{k_g,i})=0$ and $u_{q^{uu}}=1$,
    where $q^{uu}$ is an upstream queue of $q^u$ ($q^{uu}\in Q^U_{q^u}$), and
    $\hat{m}$ is the GREEN cycle index of $q^{uu}$ at $\tau_{k_g}$. In this case, $\tau_{k_g,i}'$ follows (\ref{eqn:event_der_J0}); (c) an exogenous change of $\alpha_{q^u}$ where $\tau_{k_g,i}'=0$.
This completes the state derivatives equations (\ref{eqn:state_der_Sn_alpha}) and (\ref{eqn:state_der_NEP_J0}).

    \item[4)]Event $J_{q^u,q,m}^e$ occurs at $\tau_k$ where $q^u \in Q^U_q$. As in the last case, this is an endogenous event triggered by  $r_{q^u,q,m}(\tau_k)=\Delta_{q^u,q}(\tau_k)$, with guard condition $g_k=r_{q^u,q,m}(\tau_k)-\frac{L_{q}-x_{q}(\tau_k)*l}{f_{q^u}}=0$. Similar to the analysis in the previous case, the derivative of the guard condition with respect to $\Theta_i$ is:
    \begin{multline}\label{eqn:gk_der_r}
         g_{k,i}'=r_{q^u,q,m,i}'(\tau_k^-)+f_{q^u,u,m,k-1}^r(\tau_k^-)\tau_{k,i}' + \\ \frac{l}{f_{q^u}}(x_{q,i}'(\tau_k^-)+f_{q,k-1}^x(\tau_k^-)\tau_{k,i}')=0
    \end{multline}
  The last event before $\tau_k$ that would cause the change of state $r_{q^u,q,m}$ is $G_{q^u,q,m}^e$ (end of flow burst generation). Denoting its occurrence time by $\tau_{k_e}$, we have $f_{q^u,q,m,k_e-1}^r(\tau_{k_e}^-)=0$ and $f_{q^u,q,m,k_e}^r(\tau_{k_e}^+)=1$. Therefore,
    \begin{equation}\label{eqn:gk_der_r_1}
    \begin{aligned}
    r_{q^u,q,m,i}'(\tau_k^-)&=r_{q^u,q,m,i}'(\tau_{k_e}^+)\\
    &=r_{q^u,q,m,i}'(\tau_{k_e}^-)+(0-1)\tau_{k_e,i}'\\
    &=-\tau_{k_e,i}'
    \end{aligned}
    \end{equation}
    Also by (\ref{eqn:rdynmic}) and (\ref{eqn:xdynamic}):
    \begin{equation}\label{eqn:gk_der_r_2}
        f_{q^u,q,m,k-1}^r(\tau_k^-)=1
    \end{equation}
    \begin{equation}\label{eqn:gk_der_r_3}
    f_{n,d,k-1}^x(\tau_k^-)=\alpha_q(\tau_k^-)-\beta_q(\tau_k^-)
    \end{equation}
    where the value of $\beta_q(\tau_k^-)$ follows (\ref{beta}). Then, using (\ref{eqn:gk_der_r_1}), (\ref{eqn:gk_der_r_2}) and (\ref{eqn:gk_der_r_3}) in (\ref{eqn:gk_der_r}) we finally obtain:
\begin{multline}
\label{eqn:event_der_J1}
       \tau_{k,i}'= \frac{f_{q^u}}{f_{q^u}+(\alpha_q(\tau_k^-)-\beta_q(\tau_k^-))l}*\\(\tau_{k_e,i}'-\frac{l}{f_{q^u}}{x_{q,i}}'(\tau_k^-))
\end{multline}
where $\tau_{k_e,i}'$ is the event time derivative when $G_{q^u,q,m}^e$ occurs with [$\beta_{q^u}\downarrow 0$]. This can be induced by: (a) $G2R_{q^u}$ when $\alpha_{q^u}(\tau_{k_e})>0$ or $x_{q^u}(\tau_{k_e})>0$, in which case $\tau_{k_e,i}'$ follows (\ref{eqn:event_der_zp_uparrow_theta_max}), (\ref{eqn:event_der_zp_uparrow_theta_min}), (\ref{eqn:event_der_xp_downarrow_sp}) or (\ref{eqn:event_der_xpbar_uparrow_sp});  (b) $J_{q^{uu},q^u,\hat{m}}^e$ 
where $q^{uu}$ is an upstream queue of $q^u$ ($q^{uu}\in Q^U_{q^u}$), and
    $\hat{m}$ is the GREEN cycle index of $q^{uu}$ at $\tau_{k_g}$. In this case, $\tau_{k_e,i}'$ follows (\ref{eqn:event_der_J1}); (c) an exogenous change of $\alpha_{q^u}$ where $\tau_{k_e,i}'=0$.
This completes the state derivative equation (\ref{eqn:state_der_NEP_J1}).

 \item[5)]Event $[x_q \uparrow c_q]$ occurs at $\tau_k$ and induces blocking for queue $q$. The guard condition is $g_k=x_q-c_q=0$. Similar to the calculation in case 1:
 \begin{equation}\label{eqn:event_der_start_BP}
    \tau_{k,i}'= \frac{-{x_{q,i}}'(\tau_k^-)}{\alpha_q(\tau_k^-)-\beta_q(\tau_k^-)} 
    \end{equation}
    This completes equations (\ref{eqn:state_der_start_BP_1}) and ({\ref{eqn:state_der_start_BP_2}}). 

 \item[6)]Event $[x_q \downarrow c_q]$ occurs at $\tau_k$ and releases blocking for queue $q$. Note that we cannot directly calculate $\tau_k$ through (\ref{eqn:event_time_derivative}) since $f_{q,k-1}^x(\tau_k^-)=0$. However, $[x_q \downarrow c_q]$ can only be induced by: (a) $R2G_q$, in which case $\tau_k$ follows (\ref{eqn:event_der_zp_uparrow_theta_max}), (\ref{eqn:event_der_zp_uparrow_theta_min}), (\ref{eqn:event_der_xp_downarrow_sp}) or (\ref{eqn:event_der_xpbar_uparrow_sp}); (b) [$\alpha_q\downarrow 0$], in which case $\tau_k = 0$ if $Q^U_q =\emptyset$ or $\tau_k$ follows (\ref{eqn:event_der_J1}) otherwise.
 This completes equations (\ref{eqn:state_der_end_BP_1}) and ({\ref{eqn:state_der_end_BP_2}}).  
\end{itemize}
In summary, if $G2R_q$ occurs at time $\tau_k$:
\begin{equation}
\label{eqn:event_derivative_total1}
    \tau_{k,i}' =
    \begin{cases}
    {\tau_{k_s,i}}'+\mathds{1}_{\Theta_i=\theta_p^{max}},
    & \text{if induced by [$z_p \uparrow \theta_p^{max}$]}\\ 
    
    {\tau_{k_s,i}}'+\mathds{1}_{\Theta_i=\theta_p^{min}},
    & \text{if induced by [$z_p \uparrow \theta_p^{min}$]}\\

    \frac{\mathds{1}_{\Theta_i=s_p}-x'_{q^*,i}(\tau_k^-)}{\alpha_{q^*}(\tau_k^-)-\beta_{q^*}(\tau_k^-)}, & \text{if induced by [$x_p^{max} \downarrow s_p$]}\\

       \frac{\mathds{1}_{\Theta_i=s_p}-x'_{\bar q^*,i}(\tau_k^-)}{\alpha_{\bar q^*}(\tau_k^-)},  & \text{if induced by [$x_{\bar p}^{max} \uparrow s_p$]}\\
    \end{cases} 
\end{equation}
where $q^*=\argmax_{q\in Q_p} x_q(\tau_k)$, and $\bar q^*=\argmax_{\bar q\in Q_n\backslash Q_p} x_{\bar q}(\tau_k)$ as defined in Section \ref{sec:event time derivatives}. Otherwise:
\begin{equation}
\label{eqn:event_derivative_total2}
    \tau_{k,i}' =
    \begin{cases}
           \frac{-{x_{q,i}}'(\tau_k^-)}{\alpha_q(\tau_k^-)-h_q(\tau_k^-)} , & \text{if $E_q$ occurs at $\tau_k$}\\

           \frac{f_{q^u}}{f_{q^u}-l\beta_q(\tau_k^-)} *\\
           \quad (\tau_{k_g,i}'-\frac{l}{f_{q^u}}{x_{q,i}}'(\tau_k^-)), & \text{if $J_{q^u,q,m}$ occurs at $\tau_k$}\\

            \frac{f_{q^u}}{f_{q^u}+(\alpha_q(\tau_k^-)-\beta_q(\tau_k^-))l}*\\ \quad (\tau_{k_e,i}'-\frac{l}{f_{q^u}}{x_{q,i}}'(\tau_k^-)) , & \text{if $J_{q^u,q,m}^e$ occurs at $\tau_k$}\\  

            \frac{-{x_{q,i}}'(\tau_k^-)}{\alpha_q(\tau_k^-)-\beta_q(\tau_k^-)}  , & \text{if $[x_q \uparrow c_q]$ occurs at $\tau_k$}\\  
            
    \end{cases} 
\end{equation}
Note that the event time derivative for $R2G_q$, $G_{q^u,q,m}$, $G_{q^u,q,m}^e$ and [$x_q \downarrow c_q$] can be derived from the above equation based on different cases with all events as defined in Section \ref{sec:Problem Formulation}.
 
\subsection{Cost Derivatives}
With all state derivatives obtained as shown in the previous section, we can obtain the IPA cost gradient estimator as the derivative of $L(\Theta; x(0),z(0), T)$ in (\ref{eqn:L}). Similar to the derivation provided in \cite{fleck_adaptive_2016}, the IPA estimator consisting of $dL(\Theta)/d\Theta_i$, $i=1,...,|\Theta|$ is given by
\begin{equation}\label{eqn:dL}
    \frac{dL(\Theta)}{d\Theta_i} = \frac{1}{T}\sum_{n=1}^{N}\sum_{q\in Q_n}\sum_{k=1}^{K_q} \omega_q \frac{dL_{q,k}(\Theta)}{d\Theta_i}  
\end{equation}
where
\begin{equation}\label{eqn:Lnm}
    L_{q,k}(\Theta) = \int_{\xi_{q,k}}^{\eta_{q,k}}  x_q(\Theta,t) \,dt
\end{equation}

\begin{multline}
\label{eqn:dLnm}
    \frac{dL_{q,k}(\Theta)}{d\Theta_i}=x_{q,i}'({\xi_{q,k}}^+)(t_{q,k}^{1}-\xi_{q,k})
    \\
    +x_{q,i}'({t_{q,k}^{O_{q,k}}}^+)(\eta_{q,k}-t_{q,k}^{O_{q,k}})\\
    +\sum_{o=2}^{O_{q,k}}x_{q,i}'({t_{q,k}^{o-1}}^+)(t_{q,k}^{o}-t_{q,k}^{o-1})
\end{multline}
where $O_{q,k}$ is the total number of events on the observed sample path at queue $q$ within the $k$th NEP, $t_{q,k}^{o}$ is the observed time of the $o$th event in that NEP, and $\xi_{q,k}$, $\eta_{q,k}$ are the observed occurrence times of the start and end of the $k$th NEP respectively.

It is clear from (\ref{eqn:dL}) and (\ref{eqn:dLnm}) that each IPA derivative is basically the accumulation of measurable inter-event times (timers) multiplied by a corresponding state derivative. The full information needed to evaluate this IPA estimator consists of (a) event time data, which are easy to record by directly observing events, and (b) state derivatives at those times, which are given by the simple iterative expressions derived in Sections \ref{sec:state derivative} and \ref{sec:event time derivatives}. Some state derivatives evaluated at an event time $\tau_k$ involve flow rates $\alpha_q(\tau_k)$, $h_q(\tau_k)$; however, these are \emph{only needed at specific events} 
(e.g., at the time when $S_q$ is induced by a light switching event in (\ref{eqn:state_der_Sn_light_switch})), making them easy to estimate in practice, as described in the next section. Regarding the unconstrained departure rate $h_q(t)$, considering it as a constant also makes it easy to estimate through simple offline counting methods. 

Returning to our observation in (\ref{eqn:flow_process_relationship}) that the $\gamma_{q, q_d}$ information is not needed, observe that this is indeed not involved in the derivation the IPA derivatives.

Finally, using simple online gradient-based algorithms as in (\ref{gradientopt}), we can adjust the controllable parameters to improve the overall performance and attain possibly local optima if the operating conditions do not change substantially; see also Section \ref{sec:simulation}.

\subsection{Perturbation Propagation}\label{sec:perturbation propagation}
The IPA method is able to adjust the controllable parameters automatically in an online data-driven manner. When applying it to a multi-intersection traffic system, the adjustment of the controllable parameters can synchronize the traffic lights and naturally induce a ``green wave'' effect by propagating state perturbations to downstream intersections. Such propagation is induced by the flow burst joining events and affects downstream state derivatives, thus capturing how a perturbation in a TLC parameter $\Theta_i$ causes a perturbation in the queue content $x_q(t)$. For example, if an $S_q$ event occurs induced by a $J_{q^u,q,m}$ event (where $q^u\in Q^U_q$) when $v_q=0$ (the light is RED), then by combining (\ref{eqn:state_der_Sn_alpha}) and (\ref{eqn:event_derivative_total2}), the state derivative becomes 
\begin{multline}
    x_{q,i}'(\tau_k^+)= (\beta_q(\tau_k^+)-\alpha_q(\tau_k^+))*\\
    (\tau_{k_g,i}'-\frac{l}{f_{q^u}}x_{q,i}'(\tau_k^-))
    \frac{f_{q^u}}{f_{q^u}-l\beta_q(\tau_k)}
\end{multline}
Since $\beta_q(\tau_k^+)=0$ when $v_q=0$ and $x_{q,i}'(\tau_k^-)=0$, this is reduced to  $x_{q,i}'(\tau_k^+)=-\alpha_q(\tau_k)\tau_{k_g,i}'$, where the flow rate $\alpha_q(\tau_k)$ can be traced to the upstream departure rate based on (\ref{eqn:flow_process_relationship}) and the event time derivative at $G_{q^u,q,m}$ also depends on upstream information: (a) if this is triggered by $R2G_{q^u}$, then $\tau_{k_g,i}'$ is an accumulation of light switching counts at the upstream intersection, whereas  (b) if it is triggered by $J_{q^{uu},q^u,\hat{m}}$ this again depends on the event time derivative at $G_{q^{uu},q^u,\hat{m}}$, where $q^{uu} \in Q^u_{q^u}$. Thus, recursively, $x_{q,i}'(\tau_k^+)$ is affected by all upstream states capturing how a perturbation can be propagated. A similar process takes place under other flow burst joining conditions. 

It is also worth observing that combining (\ref{eqn:state_der_En}) and (\ref{eqn:event_derivative_total2}) gives $x_{q,i}'(\tau_k^+)=0$ when event $E_q$ occurs at $\tau_k$. 
This implies that any parameter perturbation affecting $x_q(t)$ is reset to zero after such events at queue $q$. Therefore, for any perturbation at $q^u$ to propagate beyond $q$ requires that after a $J_{q^u,q,m}$ event occurs it must be followed by a $E_q$ event before a $G2R_q$ event occurs so that the vehicles from $q^u$ benefiting from a positive GREEN cycle perturbation have the chance to get through $q$ without stopping; otherwise, the perturbation at $q^u$ is ``cancelled'' by $E_q$. 
Note that such propagation only occurs through $J_{q^u,q,m}$ events, which limits computation to simple derivative updates at selected events and makes this propagation analysis scalable. This also allows us to easily track how TLC parameters affect ``green waves'' along a series of intersections.

\section{Simulation Results}\label{sec:simulation}
We use the Eclipse SUMO (Simulation of Urban MObility) simulator to build a simulation environment for traffic through a $m*n$ traffic light systems. We initially consider a $2*3$ grid shape, so that $N=6$. The length of each road segment between two neighboring intersections is set to be $300m$. For each individual intersection, we use the traffic structure shown in Fig. \ref{fig:traffic structure}: each road segment contains two queues - one for
left-turn movements and one for combined straight and right-turn movements. Four phases are designed as described in Sec.\ref{sec:Problem Formulation}: (a) [W-s],[E-s]; (b)[W-l],[E-l];(c)[S-s],[N-s]; (d)[S-l],[N-l]. Although IPA is independent of the arrival processes, we select Poisson arrival processes in SUMO to capture the traffic demand for each origin-destination (OD) pair, where the origins and destinations are different and correspond to boundary edges of the traffic network. To simplify traffic demand representation, OD pairs are grouped into four categories: row-to-row, row-to-column, column-to-row, and column-to-column. Let $\bar{\alpha} =[\bar\alpha_{rr},\bar\alpha_{rc},\bar\alpha_{cr}, \bar\alpha_{cc}]$ denote the Poisson arrival rates for each OD pair group, where subscripts indicate directionality. For example, $\bar{\alpha}_{rr}$ represents Poisson rate for OD pairs with horizontal departure and arrival edges. This abstraction allows the aggregated modeling of demand patterns between boundary edges.
Since we only need the real-time arrival flow rate at certain event times, we can estimate an instantaneous arrival rate through $\alpha_q(\tau_k)=N_q/t_w$, where $N_q$ denotes the number of vehicles joining queue $q$ during a time window of size $t_w$ before event time $\tau_k$; this is easy to detect and record in SUMO. 
We also estimate the maximum departure rate as a constant value $h_q(t)=H$ for all queues (where $H$ is determined through a separate offline analysis). 
We set $H=1.3 vehicles/s$
, $v=10m/s$, and equal weights for all flows ($w_q=1$) throughout this section. With this setting, we have performed simulation experiments to demonstrate improvements in mean waiting times, synchronization leading to ``green waves'', adaptivity, and scalability.

\subsection{Online TLC Implementation for Uncongested Traffic}
 
The ultimate goal of TLC is to operate on line, i.e., observe real-time traffic data and adaptively adjust the controllable parameters. We simulate this process by creating a single long sample path and updating IPA derivatives in (\ref{eqn:dLnm}) with every observed event occurrence (assuming some sensing capabilities for detecting events and event times). The results are accumulated and the parameters are updated every $1000$s using the data collected during the most recent time window. Two metrics are used to evaluate the performance of the TLC method: (a) Mean waiting time, which measures the time of stopping at a queue, providing an indication of delay experienced behind traffic signals; (b) Time-distance ratio, which calculates the average time taken to travel a fixed distance. This incorporates both stopping time and travel speed. A higher ratio indicates more undesirable traffic conditions associated with congestion. 
We set the initial controllable parameters for each phase of each traffic light as $\theta_p=[20,40,10]$, $p \in P_n, n=1,\ldots,6$, indicating minimum GREEN time threshold, maximum GREEN time threshold, and queue content threshold respectively. 

\begin{figure}
    \centering
    \includegraphics[width=0.95\columnwidth]{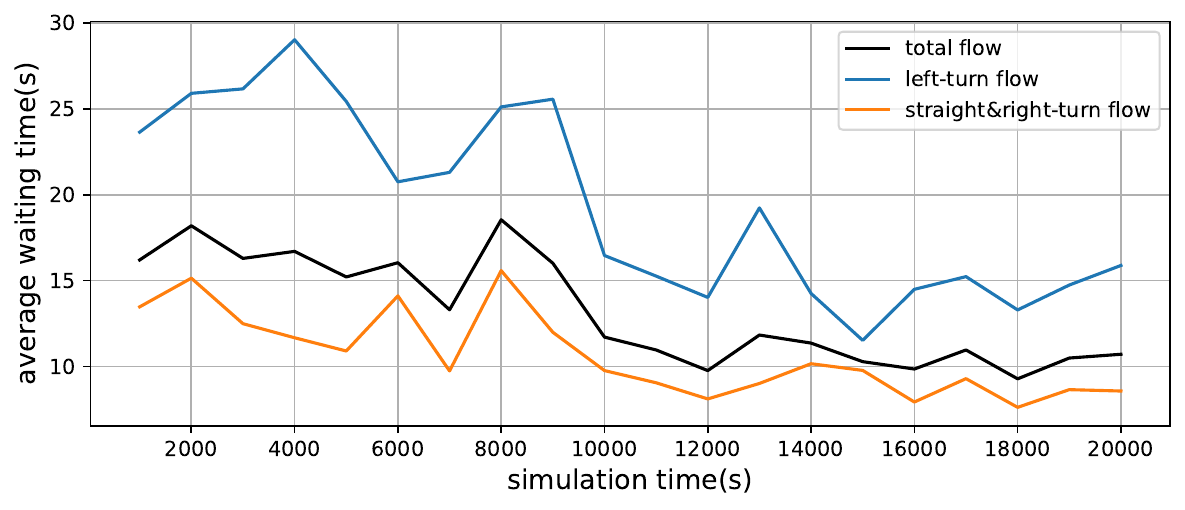}
    \caption{Sample cost trajectory for online implementation under uncongested traffic with $\bar{\alpha} =[0.02,0.01,0.02,0.01]$}
    \label{fig:performance_uncongested}
\end{figure}

We start with different relatively small $\bar{\alpha}$ values that would not lead to congested traffic, and record performances as shown in Table \ref{table:performance}. Observe that under different unbalanced traffic demands, both performance metrics show improvements from 0.75\% to 46\%. Typical sample cost trajectories for different flows under $\bar{\alpha} =[0.02,0.01,0.02,0.01]$ are shown in Fig. \ref{fig:performance_uncongested}. Observe that a left-turn queue always has a higher waiting time than a straight \& right-turn queue, while the gap becomes smaller as both values decrease.

\begin{figure}
    \centering
    \includegraphics[width=0.95\columnwidth]{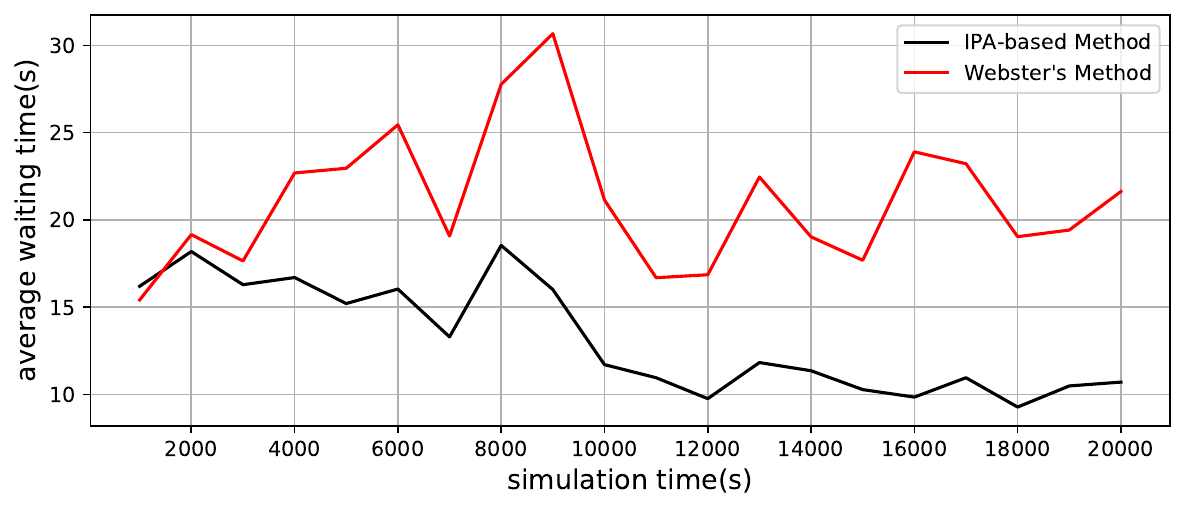}
    \caption{Sample cost trajectories comparison with Webster's Method under uncongested traffic}
    \label{fig:performance_uncongested_comparison}
\end{figure}

We also compare the performance of our method against Webster's Method \cite{reddy2016signal}, which is a rational approach for designing fixed-cycle traffic signals based on Webster's formula.
Webster's Method incorporates static information such as flow ratios and lost time to optimize predetermined cyclic plans rather than adapting in real-time. As shown in Fig. \ref{fig:performance_uncongested_comparison}, our adaptive TLC method initially has higher waiting times before optimizing the signal timings. However, it is then able to self-adjust the parameters, while the fixed cycle timing from Webster's method is prone to slight demand fluctuations, leading to consistently higher overall wait times. It should be noted that under uncongested traffic conditions, fixed-cycle policies, like Webster's method, can deliver relatively stable and acceptable performance, despite some fluctuations. However, such predetermined signal plans becomes ineffective in scenarios where congestion emerges, as demonstrated next.

\begin{table*}[]
\centering
\caption{Simulation results for different traffic demands}
\resizebox{0.8\textwidth}{!}{%
\begin{tabular}{|c|ccc|ccc|}
\hline
\multirow{2}{*}{$\bar \alpha$} &
  \multicolumn{3}{c|}{Average Waiting Time} &
  \multicolumn{3}{c|}{Time-Distance Ratio} \\ \cline{2-7} 
 &
  \multicolumn{1}{c|}{init} &
  \multicolumn{1}{c|}{opt} &
  reduction ratio &
  \multicolumn{1}{c|}{init} &
  \multicolumn{1}{c|}{opt} &
  reduction ratio \\ \hline
{[}0.02,0.01,0.01,0.01{]} &
  \multicolumn{1}{c|}{12.03} &
  \multicolumn{1}{c|}{6.42} &
  46.63\% &
  \multicolumn{1}{c|}{0.122} &
  \multicolumn{1}{c|}{0.115} &
  5.74\% \\ \hline
{[}0.02,0.02,0.01,0.01{]} &
  \multicolumn{1}{c|}{19.06} &
  \multicolumn{1}{c|}{10.81} &
  43.28\% &
  \multicolumn{1}{c|}{0.141} &
  \multicolumn{1}{c|}{0.131} &
  7.09\% \\ \hline
{[}0.02,0.01,0.02,0.01{]} &
  \multicolumn{1}{c|}{16.20} &
  \multicolumn{1}{c|}{10.71} &
  42.96\% &
  \multicolumn{1}{c|}{0.133} &
  \multicolumn{1}{c|}{0.132} &
  0.75\% \\ \hline
{[}0.02,0.01,0.01,0.02{]} & \multicolumn{1}{c|}{23.17} & \multicolumn{1}{c|}{12.37} & 46.61\% & \multicolumn{1}{c|}{0.152} & \multicolumn{1}{c|}{0.135} & 11.18\% \\ \hline
\end{tabular}%
}
\label{table:performance}
\end{table*}

\subsection{Online TLC Implementation for Congested Traffic}
We next consider scenarios where traffic demand is high enough that improper signal timing can lead to congestion and blocking. In such cases, our adaptive TLC method can detect emerging blockages and adjust the signal parameters accordingly. This allows responding to congestion in real time, which can be challenging for static traffic light schemes like Webster's Method that rely on flow ratios to determine signal cycles.
% When demand exceeds capacity, static cycles may not provide enough green time to prevent spillback. Our approach provides a responsive mechanism to tune phases and splits based on observed blocking, helping to mitigate gridlock. 

\begin{figure}
    \centering
    \includegraphics[width=0.95\columnwidth]{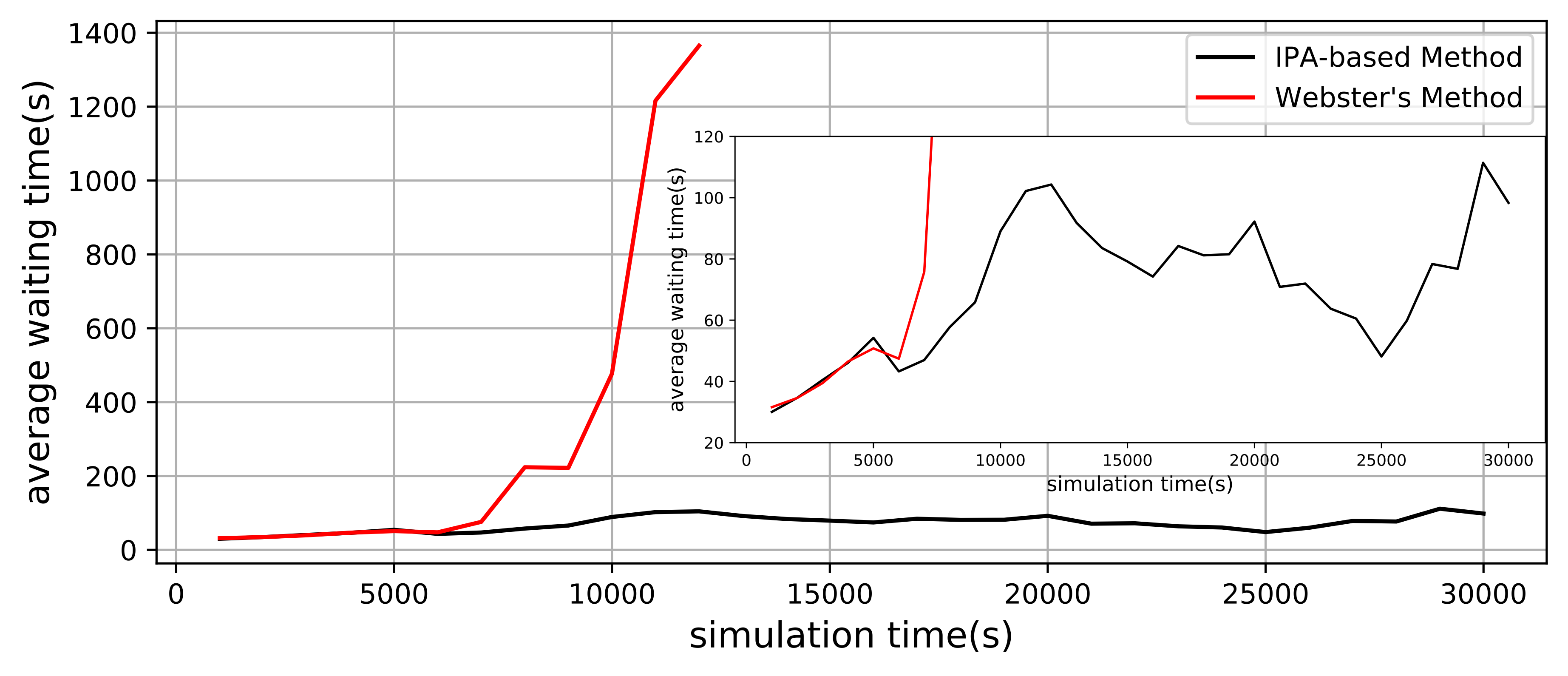}
    \caption{Sample cost trajectories comparison with Webster's Method under potential congested traffic}
    \label{fig:performance_congested_comparison}
\end{figure}

We use the same simulation setting as the uncongested case, with traffic demand set as $\bar{\alpha} =[0.02,0.02,0.02,0.011]$. A typical comparison of sample trajectories is shown in Fig. \ref{fig:performance_congested_comparison}. When traffic congestion builds up early in the simulation, our method can adapt parameters based on observed blocking to help mitigate gridlock. In contrast, the average waiting time with Webster's method increases rapidly and quickly leads to unstable conditions.

\subsection{TLC Adaptivity}
\begin{figure}
    \centering
    \includegraphics[width=0.95\columnwidth]{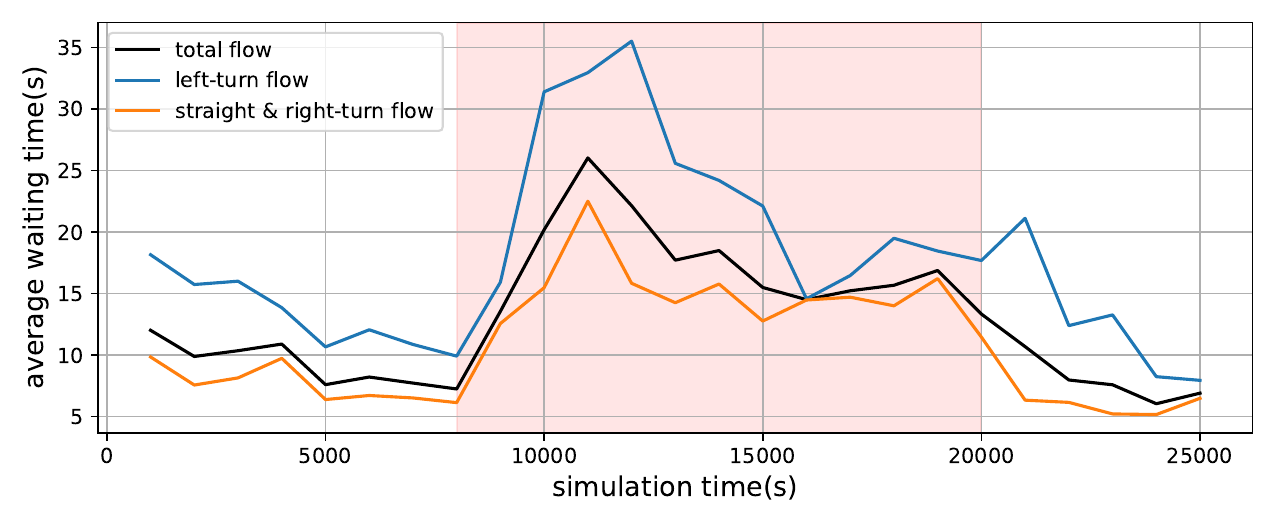}
    \caption{Sample cost trajectories with traffic demand perturbation}
    \label{fig:adaptive}
\end{figure}

Our TLC method is designed to adapt to changing traffic conditions. We illustrate this property by observing how performance changes when traffic demand is perturbed. The initial traffic demand is set through $\Bar{\alpha} = [0.01,0.01,0.01,0.015]$, with the same initial parameters as in previous cases. We add traffic perturbations by doubling the Poisson rate of the flow for row-to-column OD pairs at $t = 8000 s$ and then return to the original rate at $t =20000 s$. The cost trajectory is shown in
Fig. \ref{fig:adaptive} where the shaded area corresponds to the time interval over which traffic demand was increased.
We can see that the waiting time initially decreases due to proper parameter adjustments. When traffic demand abruptly increases, the waiting time increases as expected, since the previously optimized parameters no longer apply to the new traffic demand. Nonetheless, they gradually adjust and converge to new optimal values after several iterations, including the time interval after the traffic demand is returned to its original value. 

\subsection{TLC Scalability}
\begin{figure}
    \centering
    \includegraphics[width=0.9\columnwidth]{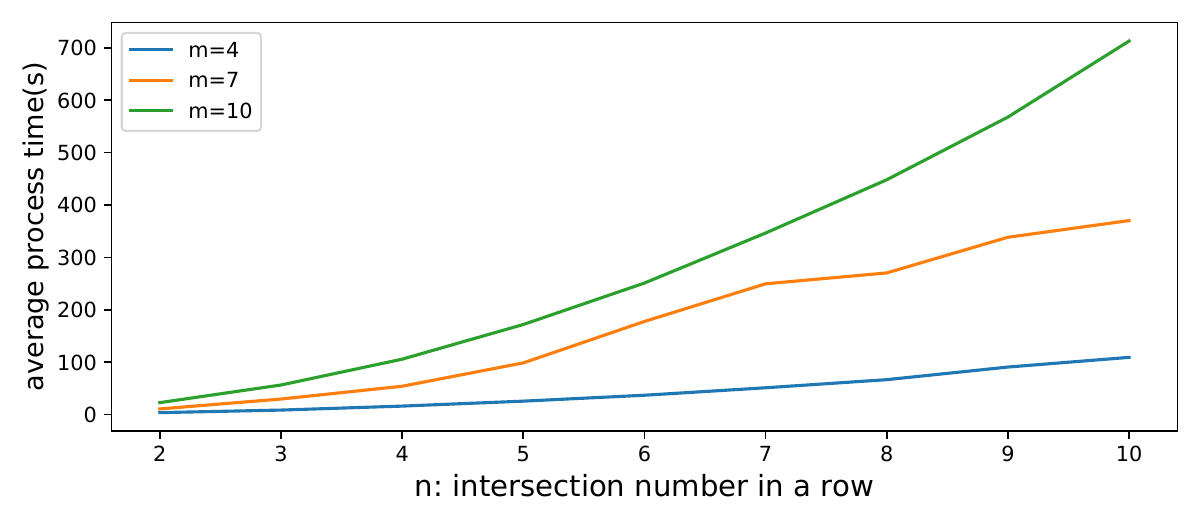}
    \caption{Scalability: CPU time as a function of the number of intersections}
    \label{fig:CPU_time}
\end{figure}

We have seen in Section \ref{section:ipa} that the IPA gradient estimation process is entirely event-driven. Therefore, the computational complexity of the TLC method is linear in the number of events. This implies that our approach scales with the number of traffic lights in a network of interconnected intersections, since the presence of a new intersections involves the addition of a similar number of events as any other and the IPA process scales accordingly.
We illustrate this property by recording the CPU time of the IPA calculations as a function of the increasing number of intersections as shown in Fig. \ref{fig:CPU_time}. With a fixed number of parallel row arteries ($m$), increasing the intersections per row from 2 to 10 results in an approximately linear increase in average IPA computation time. Furthermore, the gap in process time between proportional $m$ values increases evenly. This linear relationship indicates that the computational complexity scales well as more intersections are added. Note that in an actual traffic network the IPA calculations can be distributed over each intersection resulting in parallel processing. Given the observed linear scaling, the TLC approach can be extended to large networks while maintaining tractable computation times. 

% % \begin{figure}[t]
% % \centering
% % \begin{subfigure}{.45\textwidth}\label{fig:performance_parallel_single_dir}
% %   \centering
% %   \includegraphics[width=\textwidth]{image/performance_parallel_single_dir.pdf} 
% %   \caption{Cost trajectories}
% % \end{subfigure}
% % \begin{subfigure}{.45\textwidth}\label{fig:theta_parallel_single_dir}
% %   \centering
% %   \includegraphics[width=\textwidth]{image/theta_parallel_single_dir.pdf}  
% %   \caption{Controllable parameters trajectories}
  
% % \end{subfigure}
% % \begin{subfigure}{.45\textwidth}\label{fig:performance_parallel_single_dir_sep}
% %   \centering
% %   \includegraphics[width=\textwidth]{image/performance_parallel_single_dir_sep.pdf}  
% %   \caption{Average waiting time of each direction}
% % \end{subfigure}
% % \caption{Sample cost and parameter trajectories}
% % \label{fig:single_dir}
% % \end{figure}

\section{Conclusion and Future Work}
We have studied a TLC problem for grid-like traffic systems considering straight, left-turn, and right-turn flows. We have considered a flexible traffic structure modeling framework that provides flexibility to model heterogeneous road structures, lane configurations, traffic directions, intersection designs, and traffic signal patterns. The analysis incorporates transit delays and blocking for vehicle movements between intersections. We have used a stochastic hybrid system model and derived IPA gradient estimators of a cost metric with respect to TLC parameters (lower and upper limits to GREEN cycles, as well as thresholds on queue lengths for dynamic adjustments), accounting for delays and blocking through flow burst definitions and burst joining logic.
Based on gradient estimates, we adjust the parameters iteratively through an online gradient-based algorithm in order to improve overall performance, with the ability to automatically adapt to changing traffic conditions. Performance is measured through the mean waiting time and a time-distance ratio metric capturing speed information. Our next steps are to include additional traffic flows, such as bicycle and pedestrian traffic, similar to \cite{chen2023adaptive}, and to develop adaptive methods to automatically determine optimal gradient step sizes that can result in faster convergence.

\bibliographystyle{IEEEtran}
\bibliography{references}  %%% Uncomment this line and comment out the ``thebibliography'' section below to use the external .bib file (using bibtex) .

\end{document}